\documentclass[10pt, journal]{IEEEtran}
\usepackage{amsmath,amsfonts}
\usepackage{algorithmic}
\usepackage{array}
\usepackage{textcomp}
\usepackage{stfloats}
\usepackage{url}
\usepackage{lipsum} % Example package -- can be removed
\usepackage[utf8]{inputenc}
\usepackage{wrapfig}

\usepackage{verbatim}
\usepackage{graphicx}
\usepackage{caption,subcaption}
\usepackage{siunitx}
\usepackage{comment}
\usepackage{booktabs,floatrow}
\usepackage{multirow}

\usepackage[mathscr]{euscript}
\usepackage[algoruled,boxed,ruled,vlined]{algorithm2e}

\SetCommentSty{mycommfont}
\usepackage{pgfplots}
\usepackage{tikz}
\usepackage{setspace}
\usepackage{colortbl}
\usepackage{enumitem}
\usepackage{booktabs}
\usepackage{xcolor}
\usepackage{mathtools}
\usepackage{amsmath}
\usepackage{amssymb}
\usepackage{amsthm}
\usepackage{cite}
\usepackage{bbm}

\usepackage{float}
\restylefloat{table}

\mathtoolsset{showonlyrefs}

\usepackage{graphicx}

\usepackage{cite}
\usepackage[T1]{fontenc}
\newtheorem{theorem}{Theorem}[section]
\newtheorem{corollary}{Corollary}[theorem]
\newtheorem{lemma}[theorem]{Lemma}

\newcommand{\1}[1]{\mathbf{1}_{#1}}

\def\BibTeX{{\rm B\kern-.05em{\sc i\kern-.025em b}\kern-.08em
    T\kern-.1667em\lower.7ex\hbox{E}\kern-.125emX}}
\usepackage{balance}
\begin{document}

%%%% NOTES:
% - We need a city name for indexation purpose, even if it is redundant
% (eg: University of Atlantis, Atlantis, Atlantis)
% - \inst{} can be omitted if there is a single institute,
% or exactly one institute per author

%%%% 4. TITLE %%%%
\title{On the Response Entropy of APUFs}

\author{Vincent Dumoulin, Wenjing Rao, and Natasha Devroye \\
{\it Department of Electrical and Computer Engineering} \\
{\it University of Illinois Chicago} \\
E-mail: vdumou2, wenjing, devroye @uic.edu \\
}

\maketitle

\begin{abstract}
A Physically Unclonable Function (PUF) is a hardware security primitive used for authentication and key generation.  It takes an input bit-vector challenge and produces a single-bit response, resulting in a challenge-response pair (CRP).  
The truth table of all challenge-response pairs of each manufactured PUF should look different due to inherent manufacturing randomness, forming a digital fingerprint. 
A {\it PUF's entropy} (the entropy of all the responses, taken over the manufacturing randomness and uniformly selected challenges) has been studied before and is a challenging problem. 
Here we explore a related notion -- the  {\it response entropy}, which is the entropy of an arbitrary response given knowledge of one (and later two) other responses. 
This allows us to explore how  knowledge of some CRP(s) impacts the ability to guess another response. 

The Arbiter PUF (APUF) is a well-known PUF architecture based on accumulated delay differences between two paths. In this paper, we obtain in closed form the probability mass function of any arbitrary response given knowledge of one or two other arbitrary CRPs for the APUF architecture. This allows us to obtain the conditional response entropy and then to define and obtain the size of the entropy bins (challenge sets with the same conditional response entropy) given knowledge of one or two CRPs. 
All of these results  depend on  the probability that two different challenge vectors yield the same response, termed the response similarity of those challenges. We obtain an explicit closed form expression for this.
This probability depends on the statistical correlations induced by the PUF architecture together with the specific known and to-be-guessed challenges.
As a by-product, we also obtain the optimal (minimizing probability of error) predictor of an unknown challenge given access to one (or two) challenges and the associated predictability.

\end{abstract}

\section{Introduction}
\IEEEPARstart{P}{hysically} Unclonable Functions (PUFs) are circuits that can be integrated into chip designs to provide a low-cost digital ``fingerprint''. They show promise as hardware security primitives for Internet of Things (IoT) devices that need low-power cryptographic frameworks for authenticated communication.  PUF designs rely on randomness derived from many types of process variations, such as gate/wiring delays, designed to be the same across all chips, but inevitably differing due to random defects that occur during manufacturing \cite{7372589},\cite{MITcirca2002}. This results in an uncontrollable, unique function for each device, which is physically unclonable. 
With PUFs, the ``fingerprint'' of a device is not some stored bit-stream, but rather extracted from its unique input / output function, that may be ``challenged'' with an input vector ${\bf c} \in \{0,1\}^n$, which interacts with the random elements of a PUF instance to produce a ``response'', $R_{\bf c} \in\{\pm 1\}$. The pair $({\bf c}, R_{\bf c})$ is called a challenge-response pair (CRP). A subset of CRPs, usually randomly selected, can be used to uniquely identify or authenticate a device. There are two classes of PUFs. In weak PUFs the size of the truth table with respect to the number of random elements is small and hence must be kept secret as it is easily obtained via exhaustive evaluation; they are mostly used for key-generation in cryptography. Strong PUFs offer a huge number of CRPs, usually exponential in the number of physical random elements, and are primarily used in device authentication \cite{7372589}. How good a PUF is is often measured by their response bias, the  uniqueness, and the entropy of the PUFs, among other statistical metrics \cite{cryptoeprint:2016/320, Hori2010QuantitativeAS, rukhin2010statistical}.  The arbiter PUF (APUF), a well known strong PUF architecture, is the focus of this paper, which is formally introduced in Section \ref{sec:arch}.

A basic authentication framework for the APUF is as follows. During an enrollment phase, which takes place in a secure environment, a large set of random challenges are passed to the APUFs input and the outputs are recorded to create CRPs. This large set of CRPs (or a model of the PUF obtained via Machine Learning from this set) is then stored on the authentication server. During the authentication phase the APUF is no longer guaranteed to be in a secure environment. In order to authenticate a particular APUF according to various protocols \cite{majzoobi2012slender, rostami2014robust, yu2014noise, survey-lightweight},  a small set of challenges are used to query the APUF by the server, so as to verify the APUF's identity.

{\bf CRP correlation:} 
 The PUF community has nearly always assumed the use of random CRP sets either in  protocols or  analysis of PUF metrics, or to build multi-bit responses. 
It has not deeply studied how challenges may be correlated, how exposing one challenge (e.g. to an attacker) may impact how predictable another challenge becomes.

We focus on presenting rigorous mathematical tools to understand how much knowledge of one or two challenges reveals about the remaining challenges in an arbiter PUF. 
We propose to measure this using the {\it response entropy}, and the {\it conditional response entropy},  or entropy of a response to a challenge given (in the conditional setting) knowledge of one or two other CRPs.  This turns out to be a function of how statistically correlated the CRPs of an APUF are. We explicitly derive the response similarity $P[R_{\bf c'} = R_{\bf c}]$, which allows us to characterize the conditional response entropy and  the associated  "entropy bins" which contain all the challenges who have the same conditional entropy given knowledge of one (or two) CRPs.

{\bf Prior observations about the PUF CRP correlations:} 
Our method is based on the analytical derivation of the probability that the responses to different challenges are the same. This probability has been observed experimentally and numerically in previous work. In particular, for APUFs, figures such as \cite[Figure 6, simulated]{4681648}, \cite[Figure 12]{4700636}, comments such as those in \cite{8603753}, and partial analytical derivations (non-closed form integrals) as in \cite{Nguyen:2016:SAA:3029795.2940326} focus on the how likely the responses to two challenges differing in one bit are to be the same, and plot the 
probability that the response changes as a function of the bit flip position (or two consecutive bit positions \cite{Nguyen:2016:SAA:3029795.2940326}). However, to the best of our knowledge, this bit flip probability as an explicit closed-form function of the challenges and responses has never been derived and is one of the contributions here. We furthermore provide closed-form expressions for {\it any} two arbitrarily correlated challenges ${\bf c}$ and ${\bf c'}$ and not only ones that change in one or two bit positions. We also propose explicit algorithms enumerating all challenges with a desired response similarity to a given one or two challenges.

{\bf Prior observations about the PUF entropy:} 
In order for PUFs to be able to uniquely identify many devices, one hopes that the entropy of the PUFs manufactured in a particular architecture is large. This {\it PUF entropy}, or the entropy of all the responses (to all challenges) or a particular architecture has been studied before and is usually experimentally obtained, as it is difficult to analytically characterize.

In \cite{Rioul:2016} they asked the question of how to select challenges to maximize the entropy (of the corresponding responses) of loop PUF outputs. They used an analytical model for the loop PUF that assumes Gaussian delay elements as justified in \cite{yu2012performance}, and showed that $n$ bits of entropy (equated with the randomness or hardness of predicting a response given no other responses) may be obtained from $n$ challenges if and only if the challenges constitute a Hadamard code. 
They do not touch on the probability that two challenges result in the same response as a function of the challenges.
Later work such as \cite{Schaub2019EntropyEO} focuses on  estimating the probability distribution of certain kinds of PUFs composed of delay elements and finds the resulting Shannon entropy of the PUF is close to the max-entropy, which is asymptotically quadratic in the number of stages $n$.
\cite{JHSS1} presents a new approach for determining the min-entropy of a PUF based on convolving histograms. \cite{JHSS2} analyzes the entropy of FPGA Lookup Table-based PUFs. None of these works focus on conditional {\it response} entropy estimation given knowledge of some challenges, but rather on estimating the overall PUF entropy.  
There is also work on the statistical analysis of PUFs \cite{6800150}, and characterizing the entropy of ``strong'' PUFs \cite{Che2017AnalysisOE, 6800150} but these have been experimental and focus on the inherent qualities (bias, uniqueness and reliability) of PUFs, and do not focus on how challenges correlate the PUF responses.  

{\bf Contributions:}
for APUF, we define and obtain:

\smallskip
\noindent $\bullet$ The \emph{response similarity} between any pair of challenges ${\bf c}$ and ${\bf c'}$, $P[R_{{\bf c}'} = R_{\bf c}]$, often denoted by $p$. This response similarity is expressed as a function of the {\it similarity factor}, a function of the two challenges, and denoted as $s$. To the best of our knowledge, this is the first time that a closed form, analytical characterization of the probability that two or three challenges will produce the same response for an APUF. This underpins the Strict Avalanche Criterion (SAC) property for APUFs and yields an alternative to the Monte Carlo simulations often used in papers, e.g. \cite{stef2024_mc}  to simulate the probability that the response flips if for example one challenge input bit flips. In fact, our work provides a complete generalization of this single bit flip probability (where only one bit is flipped with respect to a base or anchor challenge) to the probability of flipping any number of bits with respect to an anchor challenge.

\smallskip
\noindent $\bullet$ The \emph{similarity bins} of any anchor challenge ${\bf c}$, $B_s(p,{\bf c})$: {for any given challenge (anchor) and a given response similarity $p$, we develop an efficient algorithm to find the set of all challenges that have the same given response similarity with respect to the anchor.}

\smallskip
\noindent $\bullet$ The \emph{response entropy} and the {\it conditional response entropy} of a challenge, conditioned on knowing one or two CRP(s). From this we  are able to compute the conditional minimum response entropy and the conditional Shannon response entropy using the closed form expression for the probability that two or three challenges will produce the same response.

\smallskip
\noindent $\bullet$ The \emph{entropy bins} of any anchor challenge or pair of challenges ${\bf c}$, $B_H(h,{\bf c})$: these are sets of challenges which all have the same response conditional entropy given ${\bf c}$. We are able to calculate the size of each entropy bin exactly. Such sets (entropy bins) can then be used to compute the expected conditional entropy of a response.

\smallskip
\noindent $\bullet$ The \emph{expected conditional entropy}: knowing the size of each entropy bin allows us to calculate the expected conditional response entropy (not the conditional PUF entropy, which is more challenging \cite{Schaub2019EntropyEO}) when knowing one or two challenges. 

\smallskip
\noindent $\bullet$  Our results yield an immediate {\bf application:} finding the optimal predictor (minimizing the probability of error) of the response to one unknown challenge ${\bf c'}$  given the response to one (and later, extended to two) known challenge(s).

\section{APUF architecture and notation}\label{sec:arch}

The APUF may be viewed as a Boolean function taking as input a challenge vector ${\bf c}:= (c_1, c_2, \cdots c_n)\in \{0,1\}^n$ and outputting a response $R_{\bf c}\in \{\pm 1\}$. This response corresponds to the output of a race resolution element, latch, or arbiter, which detects which of two racing signals 
%-- denoted by red and blue in the figures --
arrives first. The two signals traverse $n$ multiplexers in series, and in the $i$-th stage, traverse the ``parallel'' paths ($t_i, u_i$) if the challenge bit $c_i=0$, else traverse the ``crossed'' paths ($r_i, s_i$) if the challenge bit $c_i=1$. The response then is $-1$ if the upper entrance to the arbiter arrives earlier than the lower, and is $+1$ otherwise - this is modeled by whether the final accumulated delay difference $\Delta_n$ is positive or negative at the entrances of the arbiter.

In the interest of modeling this behavior analytically, in stage $i \in [1,n]$ of the APUF the delays of the four possible paths taken (parallel or crossed) are modeled as four random delay elements, $t_i, u_i, r_i$ and $s_i$, which are all i.i.d. normal random variables with mean $\mu$ and variance $\sigma^2$, denoted as $\sim {\cal N}(\mu,\sigma^2)$. they are always selected in fixed pairs, and since the response depends only on which signal arrives first, so on the relative delay difference between the two racing paths. 
The assumption that all have the same mean corresponds to the fact that the manufacturing process aims to produce identical MUXes, but due to process variation, they end up similar but not identical, with a small spread around the mean. Gaussians are good models for the inevitable manufacturing randomness, as justified in
\cite{boning2000models, yu2012performance}. Truncated Gaussians may be more suitable, but are less analytically tractable and not much better for the small variance of the manufacturing delay.

The challenge bit $c_i \in \{0, 1\}$ determines two factors: 1) which path pair ($t_i, u_i$ in parallel, or $r_i, s_i$ crossing) at stage $i$ is chosen, and 2) 
the sign of the accumulated delay difference from all the previous stages before reaching $i$. The response $R_{\bf c}$ is expressed as the sign of the accumulated delay difference at the final stage $\Delta_n$. In general, the accumulated delay $\Delta_i$ is recursively defined involving the challenge ${\bf c}$, the delay elements $r_i, s_i, t_i, u_i$ of the current stage, and the accumulated delay difference at stage $i-1$, $\Delta_{i-1}$: 
\begin{align}
R_{\bf c} = \text{sign}(\Delta_n) \in \{\pm 1 \}, & \label{eq:recursive} 
\end{align}
where $\Delta_n$ is computed recursively for $i \in [1, n]$, as% assuming $\Delta_0=0$, as:
 \begin{align}
 \Delta_i = \left\{
 \begin{array}{ll}
 +\Delta_{i-1} + t_i-u_i, &\text{when }c_i = 0 \\
 -\Delta_{i-1} + s_i-r_i, &\text{when }c_i = 1
 \end{array}
 \right. , \;\;\Delta_0=0.
\end{align}

The recursive formula is not easy for modeling and analysis, thus APUFs are frequently modeled  by transforming the $n$-bit input vector ${\bf c}$ into another $n+1$ bit vector $\Phi$, the $4n$ random variables into a vector ${\bf w}$ of size $n+1$. The response is  then expressed as a linear threshold function of ${\bf \Phi}, {\bf w}$, in a non-recursive fashion as shown in 
\cite{Tobisch:2015}: \begin{align}
 R_{\bf \Phi} = \text{sign}({\bf \Phi} \cdot {\bf w}) = \text{sign}\left( \sum_{i=1}^{n+1} \phi_i w_i\right) \in \{\pm 1 \},
 \label{eq:arbiter2}
\end{align}
where
${{\bf \Phi}}:= (\phi_1, \phi_2, \cdots \phi_{n+1}) \in \{\pm 1\}^{n+1}$ is a vector depending solely on the challenge vector ${\bf c}$, and ${\bf w}:= (w_1, w_2, \cdots w_{n+1}) \in \mathbb{R}^{n+1}$ is a vector depending only on the delay random variables, and $\cdot$ denotes the inner product:
\begin{align}
\phi_i = \left\{\begin{array}{ll}
(-1)^{\sum_{i}^{n}c_k}, & 1 \leq i \leq n, \\
 +1 & i=n+1 
\end{array}\right.\label{eq:phieq}
\end{align}
 \begin{align}
 w_i = \left\{
 \begin{array}{ll}
 (t_1 - u_1) - (s_1 - r_1), & i =1 \\ \\
 (t_{i-1}-u_{i-1}) + (s_{i-1}-r_{i-1}) \\ + (t_{i}-u_{i}) - (s_{i}-r_{i}), & 2 \leq i \leq n, \\ \\
 (t_n - u_n) + (s_n - r_n), & i = n+1 \\ 
 \end{array}
 \right.\label{eq:weq}
\end{align}
This representation eases the analysis as it separates the ${\bf \Phi}$ (dependent only on the challenge) from ${\bf w}$ (dependent only on the random delay elements), and the output is now expressed as a linear threshold function.  ML attacks against APUFs often exploit this representation.

\begin{figure}[h!]
		\centering
		\includegraphics[width=0.75\linewidth]{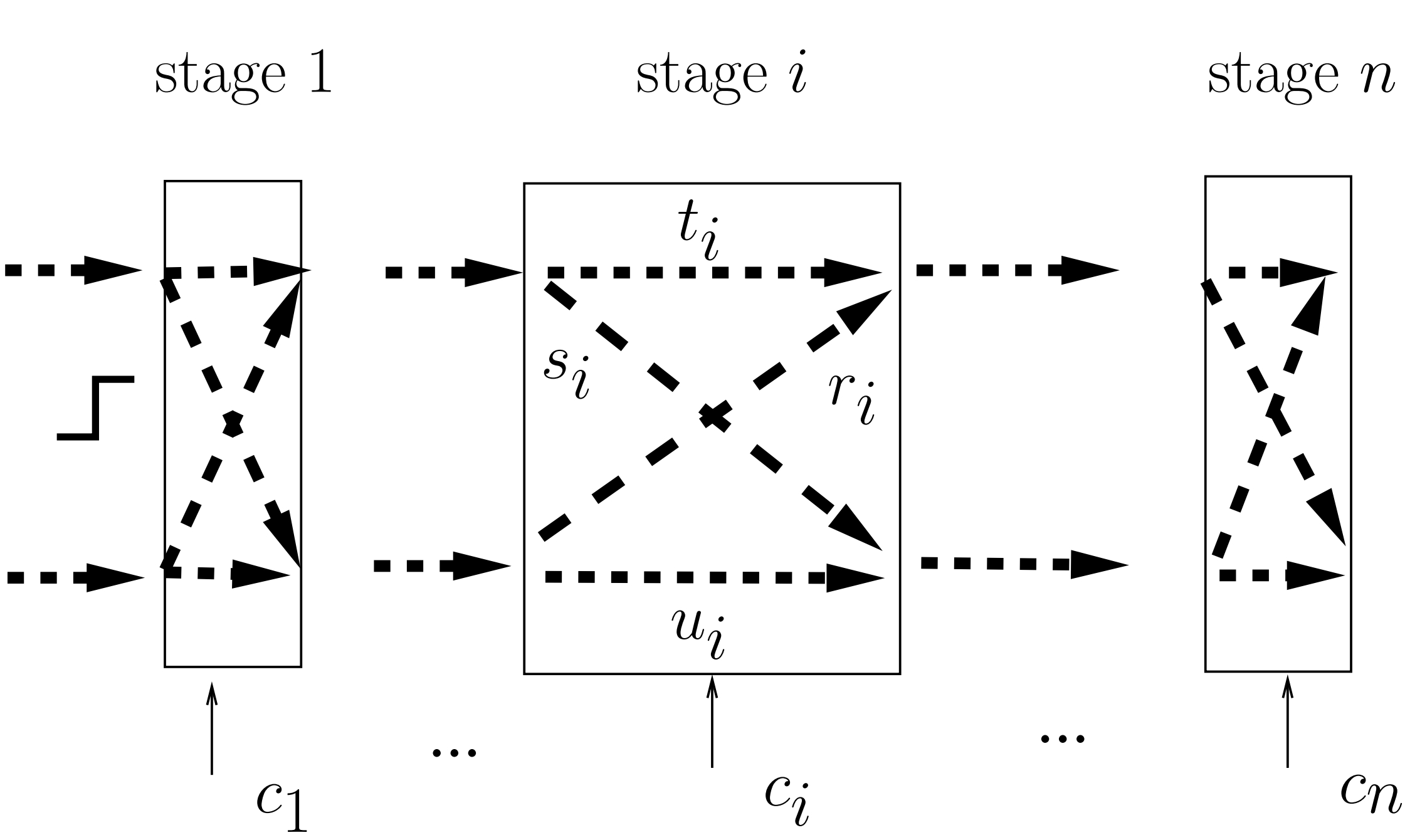}
		\caption{Illustration of the delay elements}% (this need not be the case, just for illustration).}
		\label{fig:illu_delay_elem}
\end{figure}

This transformation renders the $w_i$'s Gaussian i.i.d., all with the same variance except $w_1$ and $w_{n+1}$ (which have half the variance as the others), as outlined in the following Lemma.

\begin{lemma}
For the arbiter PUF whose output is described by Eq. \eqref{eq:arbiter2}, and using the transformation in Eq. \eqref{eq:weq}, then 
\begin{align}
w_1 \sim {\cal N}(0,4\sigma^2), \;\;w_i \sim {\cal N}(0, 8\sigma^2) \\
\text{ for } 2\leq i\leq n, \;\; w_{n+1} \sim {\cal N}(0,4\sigma^2), \label{eq:wdist}
\end{align} 
and $w_i, w_j$ are independent for $i\neq j$.
\label{lemma:w}
\end{lemma}
The transformation between the vectors ${\bf c}$ and ${\bf \Phi}$ is bijective.
For the rest of the paper we transform challenges ${\bf c}, {\bf c'}$ into ${\bf \Phi, \Phi'}$, which turns the APUF response into a linear threshold function, which is easier to deal with.

\section{Response similarity and entropy}
\label{sec:probsame}

Our technical contribution comes from the succinct derivation of the following two notions, which are then (in the following subsection) used to calculate the conditional response entropy given knowledge of another CRP.

The {\bf ``response similarity''} between ${\bf \Phi}$ and ${\bf \Phi'}$, $P[R_{{\bf \Phi}'} = R_{\bf \Phi}]$, is the probability that they produce the same response. 
The probability is taken over the random generation of the delay elements, and NOT over the fixed challenges. 
The {\bf ``similarity bins''} of an ``anchor'' challenge ${\bf \Phi}$ are sets of challenges with equal response similarity to an anchor challenge.

The main theoretical results solve these two problems for an $n$-stage arbiter PUF (with minor modifications to any other PUF that may be modeled as an $n$-stage linear threshold function as long as the randomness in the stages are independent and Gaussian distributed): 
\begin{itemize}[leftmargin=*]
	\item	{\bf Problem 1:} given challenge pair $({\bf \Phi}, {\bf \Phi'})$,
	derive their response similarity $p=P[R_{\bf \Phi} = R_{\bf \Phi'} ]$. 
    \item	{\bf Problem 2:} given a CRP $({\bf \Phi},R_{{\bf \Phi}})$, optimally predict the target response $R_{\bf \Phi'}$ of another challenge ${\bf \Phi'}$, and derive the prediction accuracy (i.e. find $P[R_{\bf \Phi'}|R_{\bf \Phi}]$)
	\item	{\bf Problem 3:} given a challenge ${\bf \Phi}$ (an ``anchor''), 
	derive all the challenges with the same response similarity $p$ to ${\bf \Phi}$, $B_s(p, {\bf \Phi})$, or the same accuracy $B_a(p, {\bf \Phi})$ (and later, the same entropy $B_H(p, {\bf \Phi})$). 
	%	list all members of similarity bin $B(p, {\bf c})$. 
\end{itemize}
{\bf Remark : } The predictor works for a particular PUF instance that is not ``abnormal'' \cite{ourpaper} i.e. there are no dominant or unusually large delay stages.

\subsection{Response similarity and accuracy}

\subsubsection{Motivational example}
\label{example}
\noindent Recall that $R_{\bf \Phi} = \text{sign}\left(\sum_{i=1}^{n+1} \phi_i w_i\right)$ and $R_{\bf \Phi'} = \text{sign}\left(\sum_{i=1}^{n+1} \phi_i' w_i\right)$, where $\phi_i = (-1)^{\sum_i^n(c_k)}$ and $\phi_i' = (-1)^{\sum_i^n(c_k')}$,
with $\phi_{n+1} = \phi_{n+1}' = 1$ for all ${\bf c, c'}$ from equations \eqref{eq:arbiter2} -- \eqref{eq:phieq}. 
We now consider the following 3 challenges, among them the likelihood of some pair resulting in the same response (over an ``average'' PUF):

\begin{align*}
 {\bf c} =& \texttt{(00000)} \quad \leftrightarrow &\Phi =& \texttt{(+1,+1,+1,+1,+1,+1)} \\
 {\bf c'} =&\texttt{(10000)} \quad \leftrightarrow &\Phi' =& \texttt{(-1,+1,+1,+1,+1,+1)} \\ 
 {\bf c''} =& \texttt{(00001)} \quad \leftrightarrow &\Phi''' =& \texttt{(-1,-1,-1,-1,-1,+1)}
\end{align*}
These challenges will lead to responses of the form (recalling that $\phi_{n+1} = 1$ always):
\begin{align*}
 R_{\bf \Phi} = &\text{sign}(+w_1+w_2+w_3+w_4+w_5+w_6) \\
 R_{\bf \Phi'} = &\text{sign}(-w_1+w_2+w_3+w_4+w_5+w_6)\\
 R_{\bf \Phi''} = &\text{sign}(-w_1-w_2-w_3-w_4-w_5+w_6)
\end{align*}
We can note the following: 1) ${\bf \Phi}$ and ${\bf \Phi'}$ are most likely to yield the same response because {their corresponding $R_{\bf \Phi}$ and $R_{\bf \Phi'}$} differ only in $w_1$;  2) ${\bf \Phi}$ and ${\bf \Phi''}$ are very likely to yield the opposite responses, because they have all the $w_i$'s in opposite signs, except for $w_6$. 
All of these observations can be confirmed by the following Theorem, where for the pair $({\bf \Phi}, {\bf \Phi'})$ the similarity factor $s = S({\bf \Phi}, {\bf \Phi'})$ is defined, and is a measure of how similar ${\bf \Phi}$ and ${\bf \Phi'}$ are.

\subsubsection{Main Theorem}
This section's main result is Theorem \ref{thm:arb2}, which shows the formal solution to Problem 1. 

All proofs of Lemmas may be found in the Appendix; the main Theorem proof is shown in the text.

\begin{theorem}[Response similarity of APUFs.] 
	
For an APUF with a pair of challenges ${\bf \Phi, \Phi'} \in \{\pm1\}^{n+1}$ their response similarity (i.e., the probability of $R_{\bf \Phi} = R_{\bf \Phi'}$) is \begin{align}
P[R_{\bf \Phi} = R_{{\bf \Phi'}}]= \frac{1}{2} + \frac{1}{\pi}\left[\arcsin\left(\frac{2{S}({\bf \Phi, \Phi'})}{n}-1\right) \right]
 \end{align}
 where 
 \begin{align}
 {S}({\bf \Phi, \Phi'}) := \frac{1}{2}\1{\phi_1=\phi_1'} + \sum_{i=2}^{n}\1{\phi_i=\phi'_i}+\frac{1}{2} \label{eq:sf}
 \end{align}	
 is the ``similarity factor'' between ${\bf \Phi}$ and ${\bf \Phi'}$. This takes on values $ \in [0,n]$ by steps of $0.5$ depending on if $\phi_1=\phi_1'$. Sometimes we drop the arguments and call it $s$ (when evaluated, as a number) for brevity. It indicates how many bits are the ``same'' in ${\bf \Phi}$ and ${\bf \Phi'}$: with $\phi_1$ and $\phi_{n+1}$ handled separately as they have half the weight as the other $w_i$'s according to the APUF-specific transformation in \eqref{eq:phieq}.
\label{thm:arb2}
\end{theorem}

{\bf Interpretation: } you can see the response similarity (i.e., the probability of $R_{\bf \Phi} = R_{\bf \Phi'}$) as the expected number of times (in \%) that the response to challenge ${\bf \Phi'}$ will be the same as the response to the anchor  ${\bf \Phi}$ where the expectation is taken across multiple PUF instances. This is equivalent to the probability (over the PUF generation process) that for a given PUF, the two challenges will have the same response.

As an example of how to use this Theorem, consider the previous example challenges ${\bf \Phi, \Phi', \Phi''}$ from Section \ref{example}. For these, $S({\bf \Phi}, {\bf \Phi'}) = 4.5, P[R_{\bf c}=R_{\bf c'}] = \frac{1}{2} + \frac{1}{\pi}\left[\arcsin\left(\frac{9}{5}-1\right) \right] \sim 0.8$, and $S({\bf \Phi}, {\bf \Phi''}) = 1/2, P[R_{\bf c}=R_{\bf c'"}] = \frac{1}{2} + \frac{1}{\pi}\left[\arcsin\left(\frac{1}{5}-1\right) \right] \sim 0.2$, aligning with the intuitive arguments before.

\subsubsection{Proof of Theorem \ref{thm:arb2}}
\noindent We can write ${\bf \Phi_1} : = (\phi_{1,1}, \phi_{1,2}, \cdots, \phi_{1,n}, 1)$ and ${\bf \Phi_2} : = (\phi_{2,1}, \phi_{2,2}, \cdots, \phi_{2,n}, 1)$. Recall that $R_{\bf \Phi_1} = \text{sign}\left(\sum_{i=1}^{n+1} \phi_{1,i} w_i\right)$ and $R_{\bf \Phi_2} = \text{sign}\left(\sum_{i=1}^{n+1} \phi_{2,i} w_i\right)$. Then, this proof follows using basics of probability and Lemma \ref{lemma:1} below. We see that 
\begin{align}
 P[R_{\bf \Phi_1} = R_{{\bf \Phi_2}}] &= 2P[\Delta_n({\bf \Phi_1}) > 0, \Delta_n({\bf \Phi_2})>0 ]\\
&\overset{(a)}{=}\frac{1}{2} + \frac{1}{\pi}\left[\arcsin \rho_{\Delta_n({\bf \Phi_1}) \Delta_n({\bf \Phi_2})}\right]\\ 
\end{align}
where $(a)$ follows by the multivariate Gaussian distribution orthant probabilities (Lemma \ref{lemma:1} below), and 
where $\rho_{AB}$ is the correlation coefficient between random variables $A$ and $B$:
\begin{equation}
  \rho_{AB} : =  \frac{E[AB]}{\sqrt{\text{Var}(A)}\sqrt{\text{Var}(B)}}.
    \label{eq:rhoAB}
\end{equation}
Let define the set $S_{ij}$ indicates the set of indices for which ${\bf \Phi_i}$ and ${\bf \Phi_j}$ are equal (excluding index $n+1$):
\begin{equation}
S_{12} : = \{i \in \{1,2,\cdots n\}: \phi_{1,i} = \phi_{2,i}\}.    \label{eq:S12}
\end{equation}

In the same way, it is possible to define the set $D_{ij}$ of indices for which ${\bf \Phi_i}$ and ${\bf \Phi_j}$ are not equal/different: % (excluding index $n+1$):
\begin{equation}
D_{12} : = \{i \in \{1,2,\cdots n\}: \phi_{1,i} \neq \phi_{2,i}\}    \label{eq:S12}
\end{equation}

So $\rho_{12} : = \rho_{\Delta_n({\bf \Phi_1}) \Delta_n({\bf \Phi_2})}$ may be calculated as 
\begin{align*}
\rho_{12}& = \frac{E[\Delta_n({\bf \Phi_1})\Delta_n({\bf \Phi_2})]}{\sqrt{\text{Var}\Delta_n({\bf \Phi_1})}\sqrt{\text{Var}\Delta_n({\bf \Phi_2})}}\\
  &= \frac{E[\left(\sum_{i\in S_{12}\cup (n+1)} \phi_i w_i \right)^2] - E[\left(\sum_{i\in D_{12}} \phi_i w_i \right)^2]}{\sqrt{4n\sigma^2}\sqrt{4n\sigma^2}} \\
  & =  \frac{\sum_{i\in S_{12}\cup (n+1)} E[|w_i|^2] - \sum_{i\in D_{12}} E[|w_i|^2]}{\sqrt{4n\sigma^2}\sqrt{4n\sigma^2}} 
    \label{eq:rhoAB}
\end{align*}
as $\text{Var}(\Delta_n({\bf \Phi_1})) = \text{Var}(\Delta_n({\bf \Phi_2}))=  4n\sigma^2$. To further evaluate the expression in Eq. \eqref{eq:rhoAB}  due to the $(\Phi, {\bf w})$ transformation/notation, we note that $E[|w_i|^2] = 4\sigma^2$ for $i\in \{2, \cdots n\}$ but that $E[|w_i|^2] = 2\sigma^2$ for $i=1, n+1$, by Lemma \ref{lemma:w}. Since index $n+1$ is always in $S_{12}$, we need to determine whether index 1 should be in $S_{12}$ or $D_{12}$. 

If $\phi_{1,1} =\phi_{2,1}$ and hence index $1$ also lies in $S_{12}$,  we have
\begin{align}
  \rho_{12}&= \frac{4\sigma^2 + 4\sigma^2(|(S_{12}| -1) - |D_{12}|)}{4n\sigma^2} = \frac{|S_{12}| - |D_{12}|}{n}.
  \end{align}
If $\phi_{1,1} = -\phi_{2,1}$ and hence index $1$ is in $D_{12}$, we have
\begin{align}
  \rho_{12}&= \frac{4\sigma^2 + 4\sigma^2(|S_{12}| - (|D_{12}|-1))}{4n\sigma^2}  = \frac{1+|S_{12}| - |D_{12}| }.{n}
\end{align}  
 Combining this with the ``similarity factor'' notation, we obtain a succinct expression for $\rho_{12}$:  
\begin{align*}
   \rho_{12} = \frac{2{S}({\bf \Phi_{1}, \Phi_{2}})}{n}-1 = \left\{\begin{array}{ll}
    \frac{2|S_{12}|}{n}-1 &\text{ if } \phi_{1,1}=\phi_{2,1} \\
    \frac{2|S_{12}|+1}{n}-1 &\text{ if } \phi_{1,1}\neq \phi_{2,1} \\
    \end{array}\right. .
\end{align*}

\begin{lemma}
Let $X\sim {\cal N}(0, \sigma^2)$,  $Y\sim {\cal N}(0,\sigma^2)$ with correlation coefficient $E[XY] = \rho_{xy}$. Then,
\begin{equation}
P[X>0, Y>0] = \frac{1}{4} + \frac{\arcsin \rho_{xy}}{2\pi}.
\label{eq:othant}
\end{equation}
\label{lemma:1}
\end{lemma}

\subsubsection{Application of Theorem \ref{thm:arb2}: }
 {\bf 
An ``optimal predictor'' $\widehat{R_{\bf \Phi_2}}(R_{\bf \Phi_1})$ 
	to predict $R_{\bf \Phi_2}$ based on the known $({\bf \Phi_1}, R_{\bf \Phi_1})$}
\label{subsubsec:opt}

\noindent One immediate application of Theorem \ref{thm:arb2} is to find the optimal predictor of the response $R_{\bf \Phi_2}$ to challenge ${\bf \Phi_2}$ once we know the response $R_{\bf \Phi_1}$ to challenge ${\bf \Phi_1}$. To find this, note that 
the conditional probability mass function
\begin{align}\label{eq:cond_proba}
&P[R_{\bf \Phi_2}|R_{\bf \Phi_1}]= \frac{P[R_{\bf \Phi_2}, R_{\bf \Phi_1}]}{P[R_{\bf \Phi_1}]}= \frac{P[R_{\bf \Phi_2}, R_{\bf \Phi_1}]}{1/2} \\
&= \left\{ \begin{array}{ll}
2 P[R_{\bf \Phi_2} =1, R_{\bf \Phi_1} =1 ]=P[R_{\bf \Phi_1} = R_{\bf \Phi_2}]  \mbox{ if } R_{\bf \Phi_2} = R_{\bf \Phi_1} \\
2 P[R_{\bf \Phi_2} =1, R_{\bf \Phi_1} =-1 ]= 1-P[R_{\bf \Phi_1} = R_{\bf \Phi_2}]  \mbox{ O/W}
\end{array} \right.  \nonumber
\end{align}
may be derived immediately from Theorem \ref{thm:arb2} and used to obtain the following {\it optimal predictor} for $R_{\bf \Phi_2}$ based on $R_{\bf \Phi_1}$, written as $\widehat{R_{\bf \Phi_2}}(R_{\bf \Phi_1})$:

\begin{corollary}
\label{cor:arb}
The optimal predictor for the response $R_{\bf \Phi_2}$ to an arbiter PUF challenged with ${\bf \Phi_2}$ given knowledge of the CRP $({\bf \Phi_1}, R_{\bf \Phi_1})$ is,
\begin{align}
&\widehat{R_{\bf \Phi_2}}(R_{\bf \Phi_1}) = \arg \max_{R_{\bf \Phi_2}\in\{\pm 1\}} P[R_{\bf \Phi_2}|R_{\bf \Phi_1}]\\ 
&= \left\{ \begin{array}{ll} 
R_{\bf \Phi_1} & \text{if }P[R_{\bf \Phi_1} = R_{\bf \Phi_2}] > 1 - P[R_{\bf \Phi_1} = R_{\bf \Phi_2}] \\
-{R_{\bf \Phi_1}} & \text{if } P[R_{\bf \Phi_1} = R_{\bf \Phi_2}] < 1 - P[R_{\bf \Phi_1} = R_{\bf \Phi_2}] \\
\end{array} \right. \label{eq:opteq}
\end{align}
where ties (when the two are exactly equal) may be broken arbitrarily. 
The prediction accuracy given by 
\begin{align}
&\max\{P[R_{\bf \Phi_1} = R_{\bf \Phi_2}], 1-P[R_{\bf \Phi_1} = R_{\bf \Phi_2}]\} \nonumber 
\label{eq:optpred}
\end{align}
with 
\begin{align*}
    P[R_{\bf \Phi_1} = R_{\bf \Phi_2}]&=\frac{1}{2} + \frac{1}{\pi}\left[\arcsin\left(\frac{2s}{n}-1\right) \right]\\
    1-P[R_{\bf \Phi_1} = R_{\bf \Phi_2}]&=\frac{1}{2} - \frac{1}{\pi}\left[\arcsin\left(\frac{2s}{n}-1\right) \right]
\end{align*}
where $s={S}({\bf \Phi_1, \Phi_2})$.
\label{cor:arbiter}
\end{corollary}

As an example application of Corollary \ref{cor:arbiter}, consider   ${\bf c}=00000$ (as this is what is physically input to the PUF we present it in this notation, but all calculations are done by transforming ${\bf c}$ to ${\bf \Phi}$ first), with $R_{\bf c} = +1$ and say we wish to predict $R_{\bf c'}$, the response to ${\bf c'} = 00110$. Since $P[R_{\bf c} = R_{\bf c'}] \sim 0.7 > (1-0.7)$, we should guess that $R_{\bf c'}$ is also equal to $+1$ and this has a probability of 0.7 of being correct.

This predictor  maximizes the probability of correctly guessing (in one guess) the value of $R_{\bf \Phi'}$ based on knowledge of $R_{\bf \Phi}$,  hence we term the predictor in \eqref{eq:opteq} the {\it optimal predictor} in this sense. 
Here, if we have one guess for $R_{\bf \Phi'}$ based on knowledge of $R_{\bf \Phi}$ this corresponds to guessing the $\widehat{R_{\bf \Phi'}}(R_{\bf \Phi})$ that maximizes $P[R_{\bf \Phi'}|R_{\bf \Phi}]$ as in \eqref{eq:opteq}.

\subsection{Entropy}
Recall the definitions of entropy of random variable (or vector) $X$ with probability mass function $P_X(x)$ taking on values $x\in {\cal X}$ and conditional entropy of random variable $X$ given random variable $Y$ (with joint distribution $P_{X,Y}(x,y)$ taking on values $x,y\in {\cal X}\times {\cal Y}$ \cite{Cover:2006}:
\begin{align}
H(X) &:= -\sum_{x\in {\cal X}} p_X(x) \log(p_X(x)) \\
H(X|Y=y) &:= -\sum_{x\in {\cal X}}p_{X|Y}(x|y) \log(p_{X|Y}(x|y)) \\
 H(X|Y) &:= - \sum_{y\in {\cal Y}} p_Y(y) H(X|Y=y)  \\
 &=  -\sum_{x\in {\cal X},y\in {\cal Y}} p_{X,Y}(x,y)\log \left( \frac{p_{X,Y}(x,y)}{p_Y(y)} \right) 
 \end{align}
We define the {\bf PUF entropy} $H(\bigcup_{\bf \Phi} R_{\bf \Phi})$ as the entropy of the vector of responses $\bigcup_{\bf \Phi} R_{\bf \Phi}$, recalling that each response is a binary random value. Note that the joint distribution $P(\bigcup_{\bf \Phi} R_{\bf \Phi})$ is hard to capture analytically as   orthant probabilities of Gaussian random vectors are generally unsolved for vectors of dimensions greater than 3, necessitating estimates or bounds on this.
The {\bf conditional PUF entropy} is the PUF entropy knowing one CRP: $H(\bigcup_{\bf \Phi} R_{\bf \Phi}|R_{\bf \Phi_1})$, which is equally hard to obtain given our inability to characterize the  Gaussian orthant probabilities beyond dimension 3.

We thus study what we call {\bf response entropy}, i.e. the entropy of one response $H(R_{\bf \Phi})$ 
and the {\bf conditional response entropy} $H(R_{\Phi_2}|R_{\Phi_1})$   given knowledge of one CRP $({\bf \Phi}, R_{\bf \Phi})$. Both the response entropy and the conditional response entropy pertain to the entropy of one binary response and hence take on values between $0$ and $1$. For the conditional response entropy $H(R_{\Phi_2}|R_{\Phi_1})$, this value will depend on how correlated $\Phi_1$ and $\Phi_2$ are, something we characterized precisely before using the response similarity.
We can use the chain rule to link all those different entropies: 
\begin{align}
    &H\left(\bigcup_{\bf \Phi} R_{\bf \Phi}\right)=\sum_{i=1}^nH(R_{{\bf \Phi}_i}|R_{{\bf \Phi}_1},\cdots,R_{{\bf \Phi}_{i-1}})\\
    &=H(R_{{\bf \Phi}_1})+H(R_{{\bf \Phi}_2}|R_{{\bf \Phi}_1})+H(R_{{\bf \Phi}_3}|R_{{\bf \Phi}_2},R_{{\bf \Phi}_1})+\cdots
\end{align}
This work calculates the first three terms (response entropies) exactly for any choice of ${\bf \Phi_1, \Phi_2, \Phi_3}$. The left term is the overall PUF entropy which is challenging to calculate.

All of the above definitions we presented the Shannon-entropy definition, but these definitions can be equivalently modified for the min-entropy. The Min-entropy of a probability mass function (or conditional probability mass function) is defined as the log of the most likely outcome. When the response $R_{\bf \Phi_1} = r_{\bf \Phi_1}$ is known, the most likely conditional entropy will be given by the probability that you guess $R_{\bf \Phi_2}$ (for a new, never before seen challenge ${\bf \Phi_2}$) correctly in one go. The min entropy is simply defined as 
\begin{equation}
 H_{\text{min}} (R_{\bf \Phi_2} | R_{\bf \Phi_1} = r_{\bf \Phi_1})
= - \log \max \{ p_s, 1-p_s \}.
\end{equation}
Min-entropy is often used when considering worst-case scenarios, ensuring that even if the distribution is not uniform, the system remains secure. Shannon entropy is more commonly used when analyzing average-case scenarios. 
In the context of hardware security primitives, min-entropy may be preferred because it helps assess the worst-case security of the PUF
when one or two CRPs have been revealed.

Without any CRP exposure, the response entropy  $P[R_\Phi=1]=P[R_\Phi=-1]=\frac{1}{2}$ so Min-response-entropy and Shannon response entropy both equal $1$ bit.
For arbitrary challenge bit vectors ${\bf \Phi_1}, {\bf \Phi_2}$ the a conditional response entropy becomes:
\begin{align}
&H(R_{\bf \Phi_2}|R_{\bf \Phi_1}=r_{\bf \Phi_1}) = -\sum_{r_{\bf \Phi_2}\in \{\pm 1 \}} P[R_{\bf \Phi_2}=r_{\bf \Phi_2}|R_{\bf \Phi_1}=r_{\bf \Phi_1}]\\
&\qquad\cdot\log_2(P[R_{\bf \Phi_2}=r_{\bf \Phi_2}|R_{\bf \Phi_1}=r_{\bf \Phi_1}])\\
& \overset{(a)}{=} - p_s \log p_s - (1-p_s)\log (1-p_s),
\label{eq:H}
\end{align}
where $(a)$ follows when we define $p_s = P[R_{\bf \Phi_2}=1|R_{\bf \Phi_1}=1]$ or $p_s = P[R_{\bf \Phi_2}=-1|R_{\bf \Phi_1}=-1]$ (depending on what value $r_{\bf \Phi_1}\in\{\pm 1\}$ takes on) and hence $1-p_s = P[R_{\bf \Phi_2}=-1|R_{\bf \Phi_1}=1]$ or $1-p_s = P[R_{\bf \Phi_2}=1|R_{\bf \Phi_1}=-1]$.

\begin{figure}[h!]
\centering
\includegraphics[width=0.8\textwidth]{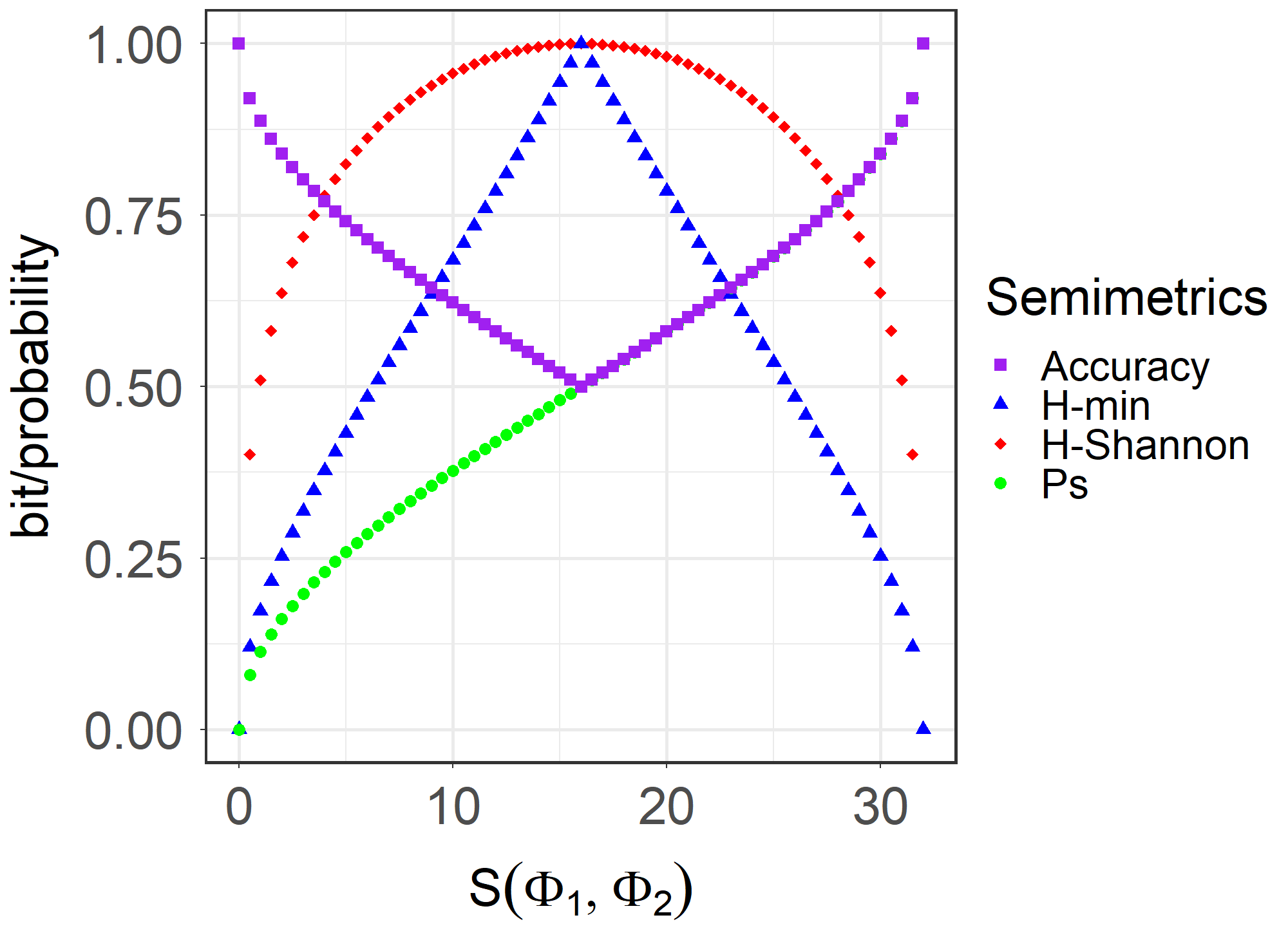} 
\caption{Accuracy, probability of same, Min-entropy and Shannon entropy of a second challenge response ${\bf \Phi_2}$ when we know 1 challenge ${\bf \Phi_1}$'s response, as a function of the ${S}({\bf \Phi_1, \Phi_2})$. Non-integer values of $S({\bf \Phi_1, \Phi_2})$ occur when $\phi_{1,1}\neq \phi_{2,1}$.}
\label{fig:entropy_know1}
\end{figure}

Figure \ref{fig:entropy_know1} illustrates the similarity factors and their corresponding response similarity $p_s = \frac{1}{2}+\frac{1}{\pi}\arcsin\left(\frac{2s}{n}-1\right)$ as a function of $s$ for a $32$-bit APUF. The x-axis shows the similarity factor $s = S({\bf \Phi_1, \Phi_2})$ between the known (say ${\bf \Phi_1}$) and the unknown (say ${\bf \Phi_2}$) challenges. The green points $p_s$ shows the response similarity ranging from 0 to 1, corresponding to the y-axis on the left.
We plot also both the Shannon and min entropies of a challenge's response when we know 1 challenge as a function of $s$. As we can see, this will depend on the correlation  $\rho_{12}$ between the challenge whose response we know, ${\bf \Phi_1}$, and the one whose response we wish to guess, ${\bf \Phi_2}$. If they are uncorrelated, i.e. $\rho_{12}=0$ (or $s$ roughly equal to $n/2$), then the response entropy remains optimal at 1 bit; the more correlated or anti-correlated they become, the more this drops, and more dramatically so for the min entropy.

\section{Similarity, accuracy and entropy Bins }

{\bf Construction of Bins:}
\noindent Given an $n$-bit challenge ${\bf \Phi_1}$ (the \textit{``anchor''}), we now show how to construct various sets containing all the challenges respecting a ``semimetric'' to the anchor. We can intuitively view a bin B of challenges as a sphere of challenges of radius $r$ that all have the same probability, a prediction accuracy or conditional entropy given the anchor challenge. This radius can be \begin{itemize}
    \item the response similarity: {\it ``similarity bins''} written $B(s, {\bf \Phi_1}) := \{{\bf \Phi_2}| {S}({\bf \Phi_1, \Phi_2}) = s\}$ or equivalently  $B(p, {\bf \Phi_1}):= \{{\bf \Phi_2}| P[R_{\bf \Phi_1} = R_{\bf \Phi_2}] = p\}$ are the  bins containing all the challenges that have the same response similarity to the anchor. For sake of the convenience, we will use the first definition $B(s, {\bf \Phi_1})$.
    \item the prediction accuracy: {\it ``accuracy bins''} written $B_a(a, {\bf \Phi_1}):= \{{\bf \Phi_2}| \max\{P[R_{\bf \Phi_1} = R_{\bf \Phi_2}], 1-P[R_{\bf \Phi_1}= R_{\bf \Phi_2}]\} = a\}$ have all the challenges which can be predicted with the same accuracy given the anchor challenge.
    \item the  entropy (Shannon or min): {\it ``entropy bins''} written $B_H(h, {\bf \Phi_1}):= \{{\bf \Phi_2}| H(R_{\bf \Phi_2}|R_{\bf \Phi_1}=r_{\bf \Phi_1}) = h\}$ all have the same conditional response entropy given the anchor challenge.
\end{itemize}
These bins $B$ can be derived using the probability $P[R_{\bf \Phi_1} = R_{\bf \Phi_2}]$ which relies  on the similarity factor. From a specific value of your ``semimetric'' you can derive the similarity factor(s) and then use algorithm \ref{alg:bin} (Appendix \ref{app:bin}). % to create these bin(s).

According to Theorem \ref{thm:arb2}, there are only $2n$ distinct response similarities thus for a given anchor challenge, all the challenges (including the anchor itself) will be partitioned into a total of $2n$ disjoint similarity bins. Intuitively, most of the challenges are in the ``uncorrelated'' bin of an anchor, i.e., $B(p\approx0.5, {\bf \Phi})$.

First, consider the trivial case $B(p=1, {\bf \Phi}) = B(s=n, {\bf \Phi})$: among all the $2^n$ challenges, the one with highest response similarity (100\%) is the anchor itself and itself only, thus $B(n, {\bf \Phi}) = \{ {\bf \Phi} \}$, and its size is 1.

Next, consider the similarity bin with the highest $p<1$: this would be $B(p, {\bf \Phi}) = B(s=n-1/2, {\bf \Phi})$, where 
$ p = \frac{1}{2}+\frac{1}{\pi}\arcsin\left(\frac{2s}{n}-1\right)$. 
For a challenge ${\bf \Phi'}$ to be in this set, it needs to satisfy $S({\bf \Phi'}, {\bf \Phi}) = n-1/2$ according to Theorem \ref{thm:arb2}. This can only be achieved by making $\phi'_1 = -\phi_1$, while keeping all other $\phi'_i = \phi_i, i \in [2, n+1]$ -- as $w_1$ contributes half the weight as the other $w_i, i \in [2, n]$, while $w_{n+1}$ always is positive (as $\phi_{n+1}=+1$). 
Thus, this will constitute $B(s=n-1/2, {\bf \Phi})$, a set with only one element.

Following the same argument, a challenge ${\bf \Phi'} \in B(s=n-1, {\bf \Phi})$ must satisfy $\phi'_1 = \phi_1$, with a single $\phi'_i = -\phi_i$ among $i \in [2, n]$. Thus the size of this bin $|B(n-1, {\bf \Phi})|= n-1$. Similarly, a challenge ${\bf \Phi'} \in B(s=n-3/2, {\bf \Phi})$ can be derived by making $\phi'_1 = -\phi_1$, and making sure only a single $\phi'_i = -\phi_i$ among $i \in [2, n]$. The size of the bin is again $n-1$.

Essentially, to obtain a challenge ${\bf \Phi'} \in B(s, {\bf \Phi})$, one needs to select $\lfloor s \rfloor$ among the anchor's $\phi_i, i\in [2, n]$ to flip their signs to form the $\phi'_i$'s. Whether $\phi'_1 = \phi_1$ or $-\phi_1$ depends on whether $s$ is an integer or not.

{\bf Size of Similarity Bins:} Algorithm \ref{alg:bin} summarizes the general method for deriving the similarity bin $B(s, {\bf \Phi})$. The complexity is linear in the size of the similarity bin to be derived. 
From Algorithm \ref{alg:bin}, we can derive the size of a similarity bin as follows: 

\begin{corollary} The size of the similarity bin $B(s, {\bf \Phi})$, for $s\in [1:n]$, for an APUF of length $n$ is 
\begin{align}\label{eq:binsize}
    |B(s, {\bf \Phi})| &= {n-1 \choose s} &\text{if }\phi_1=\phi_1'\\
    &= {n-1 \choose \lfloor s-1 \rfloor}\ &\text{if }\phi_1\neq\phi_1'
\end{align}
\label{cor:binsize}
\end{corollary}

\begin{figure}[h!]
\centering
\includegraphics[width=0.8\textwidth]{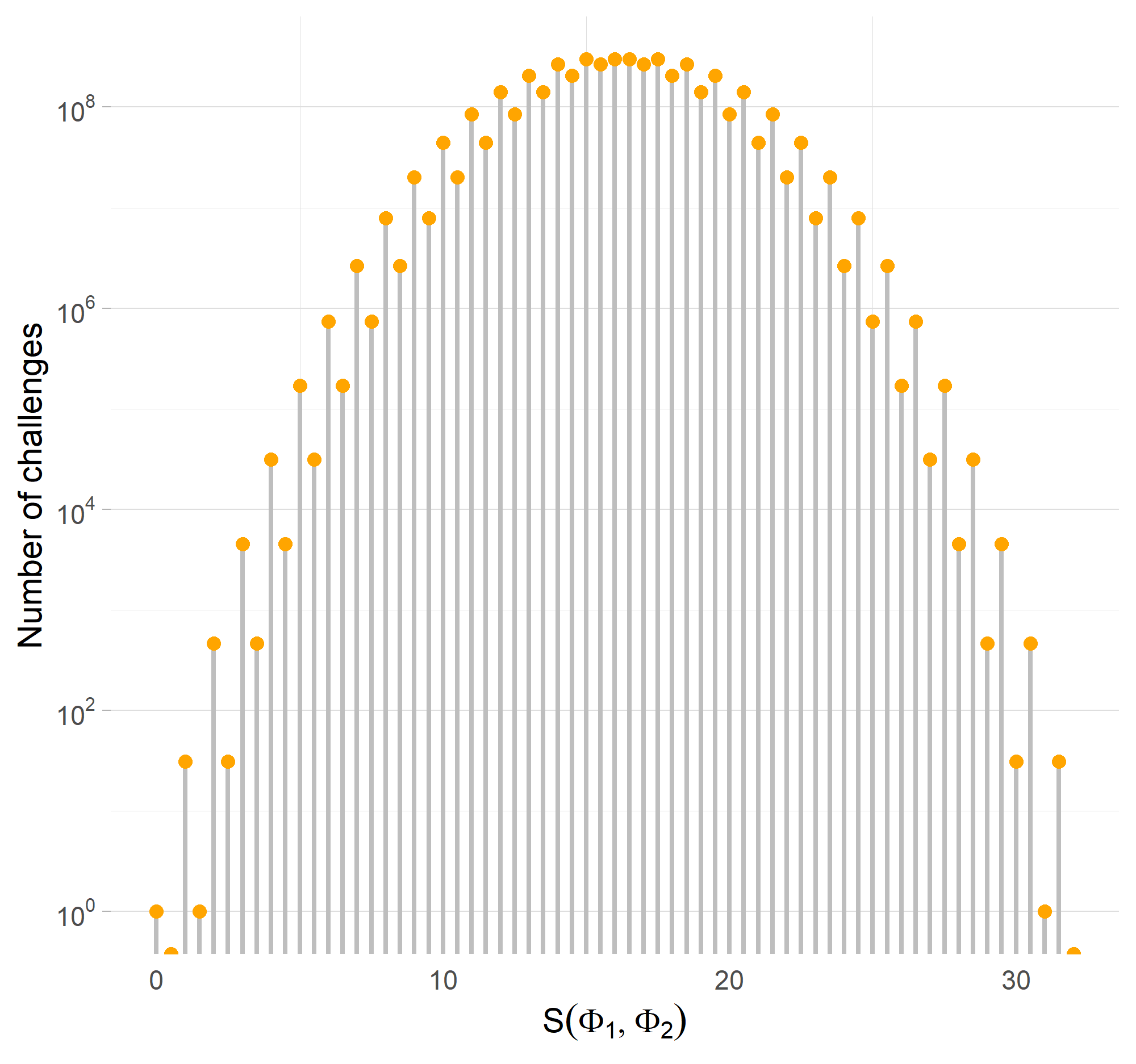}
\caption{Similarity bin sizes (number of challenges) in function of ${S}({\bf \Phi_1, \Phi_2})$. Not integer values of S means that $\phi_{1,1}\neq \phi_{2,1}$.}
\label{fig:density_know1}
\end{figure}

Figure \ref{fig:density_know1} shows the size of each similarity bin, $|B(s, {\bf \Phi})|$, obtained from Corollary \ref{cor:binsize} on the right y-axis.

{\bf Expected conditional entropy:} 
\noindent From Figures \ref{fig:entropy_know1} and \ref{fig:density_know1} we have all the information needed to compute the expected conditional response entropy or accuracy of a response given one CRP. We know the specific value of the response entropy for each possible known challenge as well as the exact number of challenges in an entropy bin,  i.e the exact number of challenges such that $H(R_{\bf \Phi'}|R_{\bf \Phi}=r_{\bf \Phi})=h$ since $B(h, {\bf \Phi})$ is the union of two disjoints $B(s, {\bf \Phi})$. Defining $h_{s,{\bf \Phi}}$ as $h_{s,{\bf \Phi}}=H(R_{\bf \Phi'}|R_{\bf \Phi}=r_{\bf \Phi})$ with ${\bf \Phi}'\in B(s,{\bf \Phi})$, the expected response entropy becomes the weighted average of the response entropies, weighted by the number of challenges in each bin as below:
\begin{equation}
    \Bar{H}(R_{\bf \Phi'}|R_{\bf \Phi}=r_{\bf \Phi})=\frac{1}{2^n-1}\sum_{s\in\{0,\frac{1}{2},1,\cdots, n-\frac{1}{2}\}} h_{s,{\bf \Phi}}\cdot |B(s, {\bf \Phi})|
\end{equation}
The expected accuracy (of estimating a next challenge given knowledge of one CRP) can be similarly calculated as a weighted sum. The expected conditional min and Shannon entropies and the expected conditional accuracy are calculated in Table \ref{tab:expectedk1p1} for different PUF lengths. Note that this does not depend on the variance $\sigma^2$ of the delay element generation process. 
\begin{table}[h!]
\begin{center}
\begin{tabular}{ |c|c|c|c| } 
\hline
&n=32 & n=64 & n=128 \\
\hline
 H-min& 0.8747 & 0.9112 & 0.9369 \\ 
H-Shannon&0.9900 &0.9952 & 0.9977 \\ 
Accuracy&0.5465 &0.5323 & 0.5226 \\ 
\hline
\end{tabular}
\end{center}
\caption{Expected conditional entropy and expected accuracy of a challenge's response knowing one CRP for APUFs of different length (number of stages n).}
\label{tab:expectedk1p1}
\end{table}

\section{Scalability to more known CRPs}
\label{sec:scalability}

\noindent All conditional response entropy and optimal predictor results so far have assumed a single anchor: i.e.,  {\it one} CRP $({\bf \Phi}, R_{\bf \Phi})$ to be known. The natural next question is how to obtain optimal predictors and conditional response entropies with {\it multiple} anchors. 
This is based on finding the conditional probability mass function, which, at its core, depends on the availability of closed form expressions for the orthant probabilities of jointly Gaussian random variables. While this is known for Gaussian vectors of dimensions 2 (useful for one known, one target challenge) and 3 (useful for two known, and one target challenge), it remains unknown for larger dimensions \cite{10.2307/2239206}, and hence there is little hope for {\it closed form} solutions for higher dimensions, i.e. optimally predicting the response to a target challenge given knowledge of 3 or more known CRPs. Numerical approximations are left for future work.

\subsection{Optimal predictor}
%\nrd{Naive way might be to let each known one predict one and then if disagree take one that is most correlated. Turns out this is exactly what the optimal predictor does: it takes the most correlated of the two and uses that. mention later on when get there. If know 3 and predict one we can continue on, but unknown whether optimal and cannot easily obtain as would need the Gaussian quadrant probabilities, which is an open question in statistics.}
When $({\bf \Phi_1}, R_{\bf \Phi_1})$ and $({\bf \Phi_2}, R_{\bf \Phi_2})$ are known and we wish to predict the response $R_{\bf \Phi_3}$ to a third challenge ${\bf \Phi_3}$, one naive approach may be to use our previous know-one-predict-one predictor to predict $R_{\bf \Phi_3}$ based on the response of the ``most correlated'' challenge  to ${\bf \Phi_3}$, i.e. based on the response predicted by either ${\bf \Phi_1}$ or ${\bf \Phi_2}$ 
%$({\bf \Phi_1}, R_{\bf \Phi_1})$, do this again to predict $R_{\bf \Phi_3}$ based on  $({\bf \Phi_2}, R_{\bf \Phi_2})$. If they agree, that is your final prediction; if they disagree then take the predictor of the challenge ${\bf \Phi_1}$ or ${\bf \Phi_2}$ that is most correlated to ${\bf \Phi_3}$. 
We will show that this is in fact provably optimal. This naive strategy can be extended to knowing more than 2 CRPs to predict another, but we are unable to prove optimality, which essentially stems from Gaussian orthant probabilities being unknown for more than 3 dimensions.

For knowing two CRPs and predicting a third, the optimal predictor takes on the following form \cite{Massey:guessing}:
\begin{equation}
\widehat{R_{\bf \Phi_3}} = \arg \max P(R_{\bf \Phi_3}|R_{\bf \Phi_1}, R_{\bf \Phi_2}).
\label{eq:opt1from1}
\end{equation}
Hence, to obtain this optimal predictor, we need to obtain the conditional probability mass functions $P(R_{\bf \Phi_3}|R_{\bf \Phi_1}, R_{\bf \Phi_2})$. This involves finding the eight values
\begin{equation}
    P(R_{\bf \Phi_3}=r_3|R_{\bf \Phi_1}=r_1, R_{\bf \Phi_2}=r_2), \;\;\;\; r_i \in \{\pm 1\}.
    \label{eq:2to1compact}
\end{equation}
We only need the four values $P(R_{\bf \Phi_3}=-1|R_{\bf \Phi_1}=r_1, R_{\bf \Phi_2}=r_2)$ from which we can find the others as $ P(R_{\bf \Phi_3}=1|R_{\bf \Phi_1}=r_1, R_{\bf \Phi_2}=r_2) = 1- P(R_{\bf \Phi_3}=-1|R_{\bf \Phi_1}=r_1, R_{\bf \Phi_2}=r_2) $.
By definition, we have that
\begin{align}
 &P(R_{\bf \Phi_3}=r_3|R_{\bf \Phi_1}=r_1, R_{\bf \Phi_2}=r_2)\\ &= \frac{P(R_{\bf \Phi_3}=r_3, R_{\bf \Phi_1}=r_1, R_{\bf \Phi_2}=r_2)}{P(R_{\bf \Phi_1}=r_1, R_{\bf \Phi_2}=r_2)}.
 \label{eq:1from2}
 \end{align}
Recall that $R_{\bf \Phi} = \text{sign}(\Delta_n({\bf \Phi}))$, and that $\Delta_n$ is a Gaussian random variable, as it is the sum of Gaussian random variables.  To obtain \eqref{eq:1from2} we thus need only calculate $P(R_{\bf \Phi_3}, R_{\bf \Phi_1}, R_{\bf \Phi_2})$ and $P(R_{\bf \Phi_1}, R_{\bf \Phi_2})$. These may both be obtained by noting that since $R_{\bf \Phi} = \text{sign}(\Delta_n({\bf \Phi}))$ are signs of zero mean, equal variance Gaussian random variables, these probabilities amount to the orthant probabilities of jointly Gaussian random variables, i.e.
\begin{align}
&P(R_{\bf \Phi_3}= 1|R_{\bf \Phi_1}=1, R_{\bf \Phi_2}=1) \\
& = \frac{P(R_{\bf \Phi_3}=1, R_{\bf \Phi_2}=1, R_{\bf \Phi_1}=1)}{P(R_{\bf \Phi_1}=1, R_{\bf \Phi_2}=1)} \\
& = \frac{P(\Delta_n({\bf \Phi_3})>0, \Delta_n({\bf \Phi_2})>0, \Delta_n({\bf \Phi_1})>0)}{P(\Delta_n({\bf \Phi_1})>0, \Delta_n({\bf \Phi_2})>0))}. \label{eq:last}
\end{align}
If $X,Y,Z$ are zero mean Gaussian random variables, then
\begin{align}
&P(X>0,Y>0,Z>0)\\
&= \frac{1}{8}+\frac{1}{4\pi}\left[\arcsin\rho_{XY} + \arcsin\rho_{XZ} + \arcsin \rho_{XZ} \right] \\
&P(X>0, Y>0) = \frac{1}{4}+\frac{1}{2\pi}\left[\arcsin\rho_{XY} \right] \\
&P(X>0|Y>0,Z>0) \\
&= \frac{1}{2}\left[1+ \frac{\arcsin\rho_{XY} + \arcsin\rho_{XZ}}{\frac{\pi}{2}+\arcsin\rho_{YZ}}\right],
\end{align}

Letting $\rho_{ij}$ be the correlation coefficient between $\Delta_n({\bf \Phi_i})$ and $\Delta_n({\bf \Phi_j})$ (which we recall are Gaussian random variables)  we obtain the following:

\begin{theorem}
     For $r_1, r_2\in \{\pm 1\}$, 
     \begin{align}\label{eq:main_k2_p1}
        &P(R_{\bf \Phi_3}= 1|R_{\bf \Phi_1}=r_1, R_{\bf \Phi_2}=r_2) \\
        &=  \frac{1}{2}\left[1+ \frac{r_1\arcsin\rho_{13} + r_2\arcsin\rho_{23}}{\frac{\pi}{2}+r_1r_2\arcsin\rho_{12}}\right]    
    \end{align}
    and we can obtain $   P(R_{\bf \Phi_3}= -1|R_{\bf \Phi_1}=r_1, R_{\bf \Phi_2}=r_2) = 1-    P(R_{\bf \Phi_3}= 1|R_{\bf \Phi_1}=r_1, R_{\bf \Phi_2}=r_2)$.
    \label{thm:k2p1}
\end{theorem}
 For clarity, we detail the quantities needed:
    \begin{align}
       | S_{ij} | &: = \# \text{ indices }l \in \{1,2,\cdots n\} \text{ for which } \phi_{i,l} = \phi_{j,l}   \label{eq:S12again} \\
    \rho_{ij}   & = \left\{\begin{array}{ll}
    \frac{2|S_{ij}|}{n}-1 &\text{ if } \phi_{i,1}=\phi_{j,1} \\
    \frac{2|S_{ij}|+1}{n}-1 &\text{ if } \phi_{i,1}\neq \phi_{j,1} \\
        \end{array}\right. .
    \end{align}

From the above equation \eqref{eq:main_k2_p1} we notice a few things: 1) first, the prediction accuracy depends on the actual response values $r_1, r_2$ and how they interact with the correlation coefficients. For example, for good prediction accuracy, you want the second term inside the brackets to be close to $1$ or $-1$ (and hence the overall prediction being close to either 1 or 0). Poor accuracy corresponds to the second term being close to $0$ (and hence the overall prediction being around $0.5$). For good accuracy you want $r_1\arcsin\rho_{13}$ and $r_2\arcsin\rho_{23}$ to  align or point in the same direction, both adding to a positive $+1$ or a negative $-1$, then it will be easy to predict the third $r_3$ as it will likely be in the same direction as the others.  If the two challenge responses and correlation coefficients contradict each other, i.e. if $r_1\arcsin \rho_{13} = -r_2\arcsin \rho_{23}$ then the accuracy will be poor. The denominator also matters: if it becomes close to $0$, or if $r_1 r_2\arcsin \rho_{12}$ is close to $-\pi/2$ (could happen if $r_1 = r_2 = +1$ but $\arcsin \rho_{12} = -\pi/2$ which means that $\rho_{12} = -1$ and the challenges are statistically anti-correlated yet produced the same sign -- virtually impossible), then the prediction accuracy is close to $\frac{1}{2}$.

The optimal predictor of $R_{\bf \Phi_3}$ given knowledge of $R_{\bf \Phi_2}$ and $R_{\bf \Phi_1}$ first calculates $\rho_{12}, \rho_{13}$ and $\rho_{23}$. From this, and   $R_{\bf \Phi_2}$ and $R_{\bf \Phi_1}$  we select $R_{\bf \Phi_3}$ as:
\[ \widehat{R_{\bf \Phi_3}} = \arg \max_{R_{\bf \Phi_3} \in \{\pm 1\}} P(R_{\bf \Phi_3}|R_{\bf \Phi_1}, R_{\bf \Phi_2}) .\]
The optimal prediction accuracy is then given by
\begin{align}
  &\text{Prediction accuracy}= \max\bigl\{ P[R_{\bf \Phi_3}=1|R_{\bf \Phi_1}=r_1, R_{\bf \Phi_2}=r_2],\nonumber \\
 & \;\;\;\;\;\; P[R_{\bf \Phi_3}=-1|R_{\bf \Phi_1}=r_1, R_{\bf \Phi_2}=r_2]\bigr\} \label{eq:pred2to1}
\end{align}
which may be re-written as a value  $A$ as:
\begin{align}
    A = \max\Bigl\{& \frac{1}{2}\left[1+\frac{r_1\arcsin{\rho_{13}}+r_2\arcsin{\rho_{23}}}{\pi/2+r_1r_2\arcsin{\rho_{12}}}\right],\\ & 1-\frac{1}{2}\left[1+\frac{r_1\arcsin{\rho_{13}}+r_2\arcsin{\rho_{23}}}{\pi/2+r_1r_2\arcsin{\rho_{12}}}\right]\Bigr\} .
\end{align}
This shows that the optimal predictor may be obtained in closed form for the arbiter PUF when two CRPs are known. In fact the optimal predictor has a simple form: if you know 2 challenges and want to predict a third, simply select the closest challenge to the anchor (the most correlated) and use the know-one-predict-one predictor. In other words, the response of the third challenge is the response of closest of the two challenges times the sign of its correlation:
\begin{align*}
   & P(R_{\bf \Phi_3}= 1|R_{\bf \Phi_1}=r_1, R_{\bf \Phi_2}=r_2) \gtrless 0.5\\
 &\Leftrightarrow    \frac{1}{2}\left[1+ \frac{r_1\arcsin\rho_{13} + r_2\arcsin\rho_{23}}{\frac{\pi}{2}+r_1r_2\arcsin\rho_{12}}\right]\gtrless0.5\\
  &\Leftrightarrow   \frac{r_1\arcsin\rho_{13} + r_2\arcsin\rho_{23}}{c}\gtrless0\\
   &\Leftrightarrow  r_1\arcsin\rho_{13}+ r_2\arcsin\rho_{23}\gtrless 0
      \end{align*}
   \begin{align*}
   &\Leftrightarrow r_1\rho_{13}+ r_2\rho_{23}\gtrless 0\\
   &\Leftrightarrow R_{\bf \Phi_3}= r_i \cdot \text{sign}(\rho_{i3}) \, \text{with } i={\arg\max}{\{|\rho_{13}|,|\rho_{23}|}\}  
\end{align*}

\subsection{Entropy}
The conditional response entropy given two CRPs also depends only on the conditional probability mass functions, this is easily obtained once we have the conditional distributions, as:
\begin{align*}
&H(R_{\bf \Phi_3}|R_{\bf \Phi_1}=r_{\bf \Phi_1}, R_{\bf \Phi_2} =r_{\bf \Phi_2})\\
&= -\sum_{r_{\bf \Phi_3}\in \{0,1\}} P(R_{\bf \Phi_3}=r_{\bf \Phi_3}|R_{\bf \Phi_1}=r_{\bf \Phi_1},  R_{\bf \Phi_2}=r_{\bf \Phi_2}) \cdot \\
&\qquad\log_2(P(R_{\bf \Phi_3}=r_{\bf \Phi_3}|R_{\bf \Phi_1}=r_{\bf \Phi_1},  R_{\bf \Phi_2}=r_{\bf \Phi_2})).
\end{align*}

\begin{figure*}[h!]\captionsetup[subfigure]{font=scriptsize}
\centering
\begin{subfigure}{0.25\textwidth}
  \includegraphics[width=\linewidth]{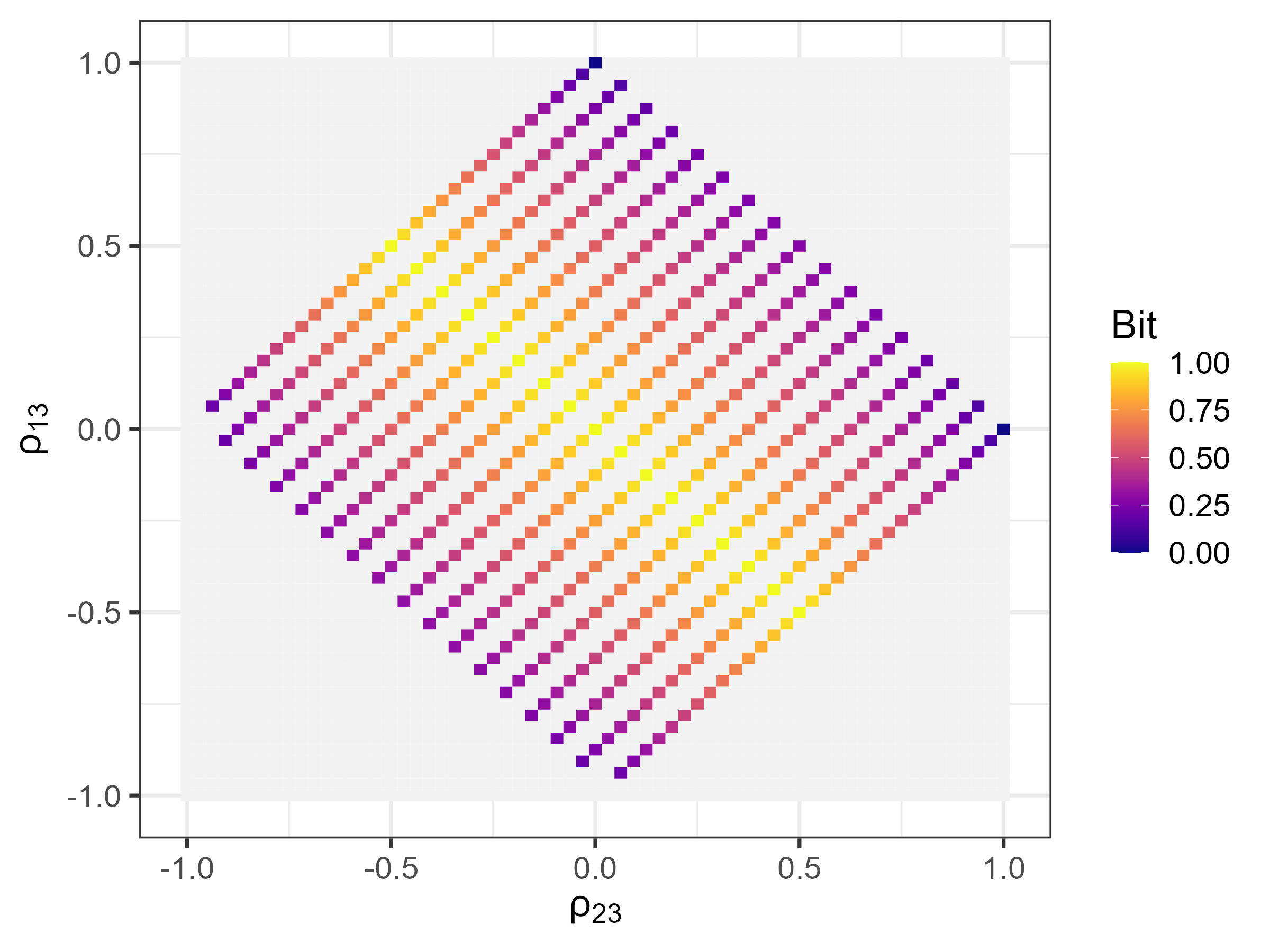}
  \caption{Min-Entropy : $r_1=r_2$ and $\rho_{12}=0$}
  \label{fig:1}
\end{subfigure}\hfil % <-- added
\begin{subfigure}{0.25\textwidth}
  \includegraphics[width=\linewidth]{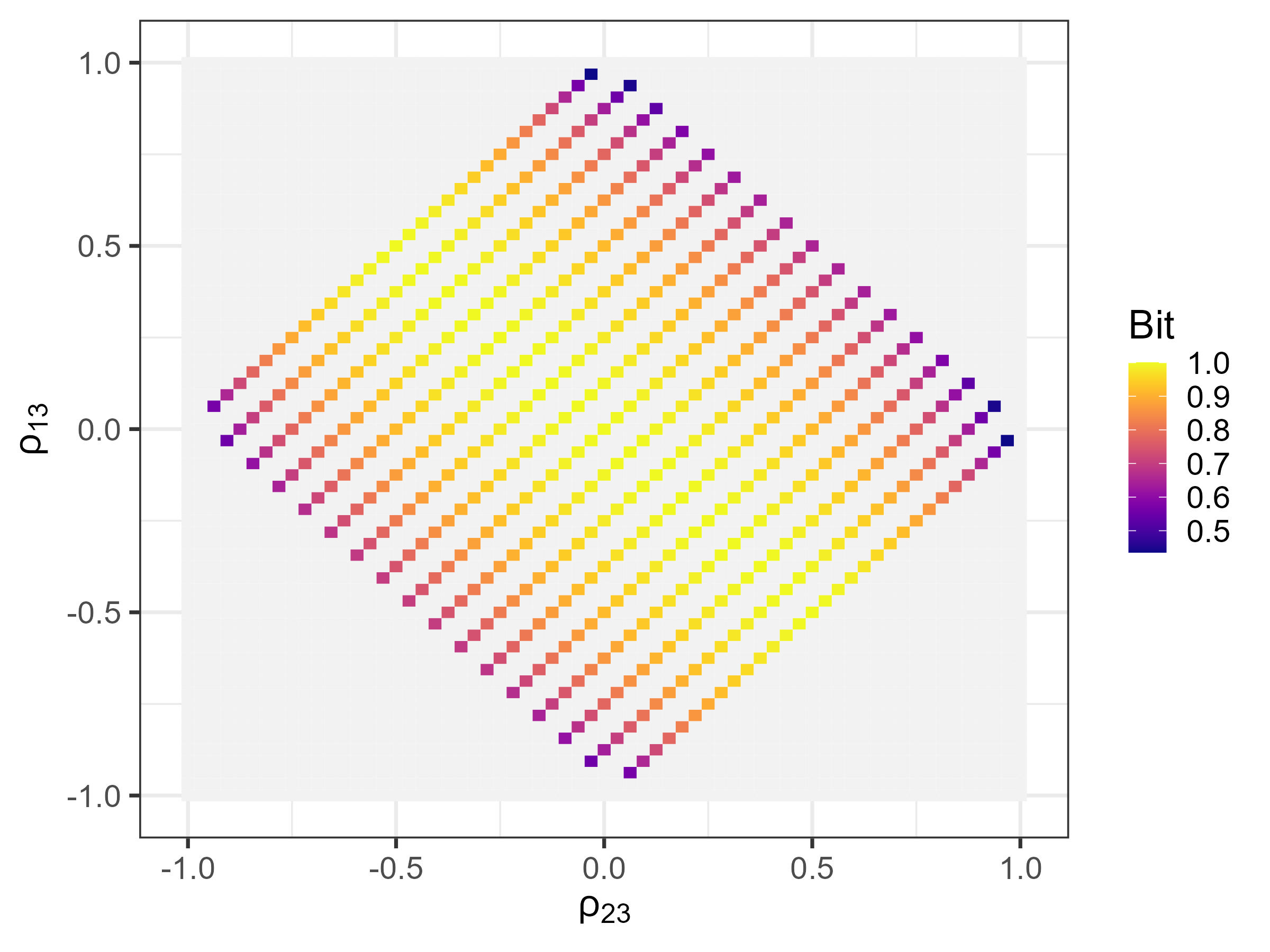}
  \caption{S-Entropy : $r_1=r_2$ and $\rho_{12}=0$}
  \label{fig:2}
\end{subfigure}\hfil % <-- added
\begin{subfigure}{0.25\textwidth}
  \includegraphics[width=\linewidth]{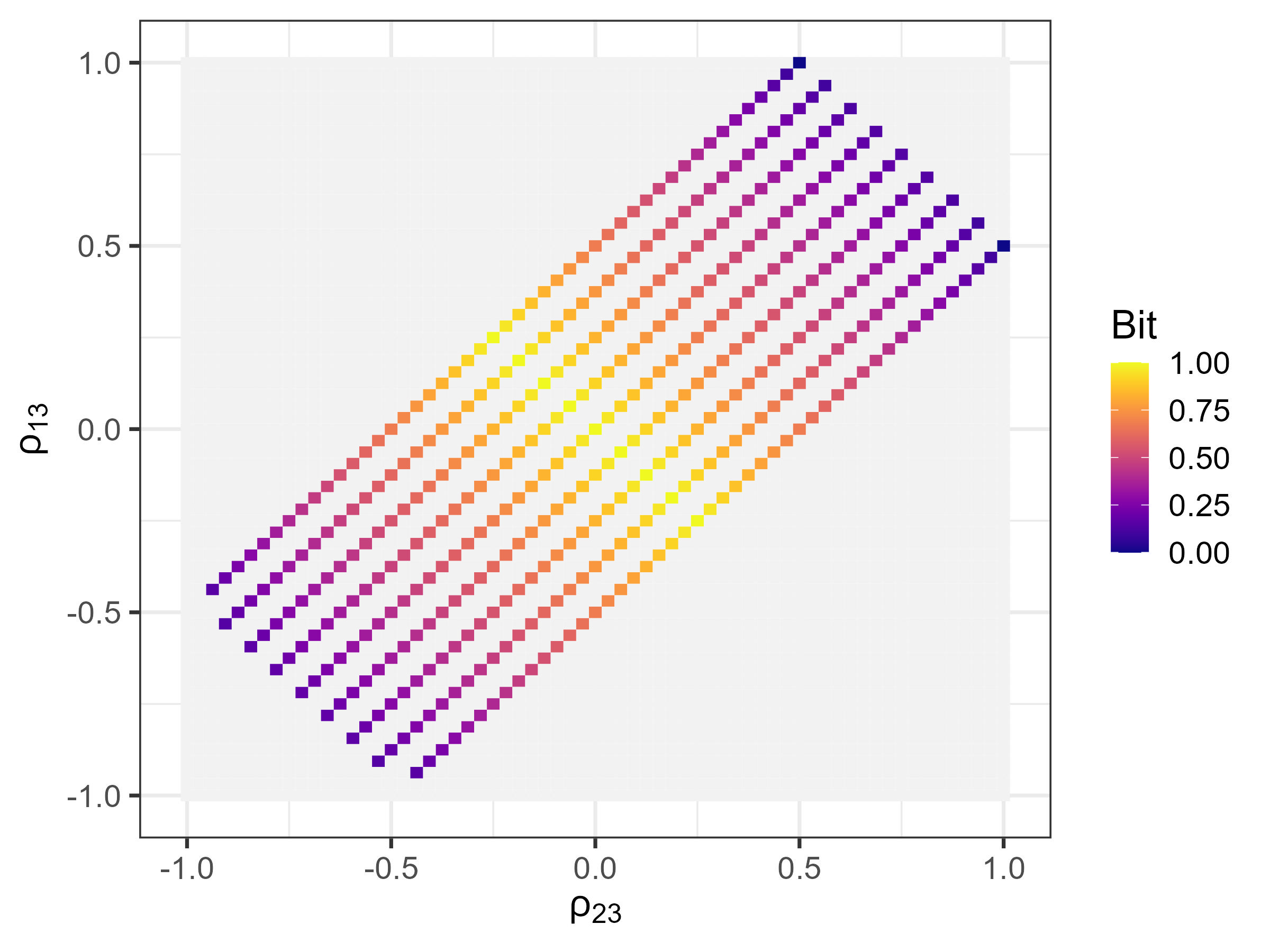}
  \caption{Min-Entropy : $r_1=r_2$ and $\rho_{12}=0.5$}
  \label{fig:3}
\end{subfigure}\hfil
\begin{subfigure}{0.25\textwidth}
  \includegraphics[width=\linewidth]{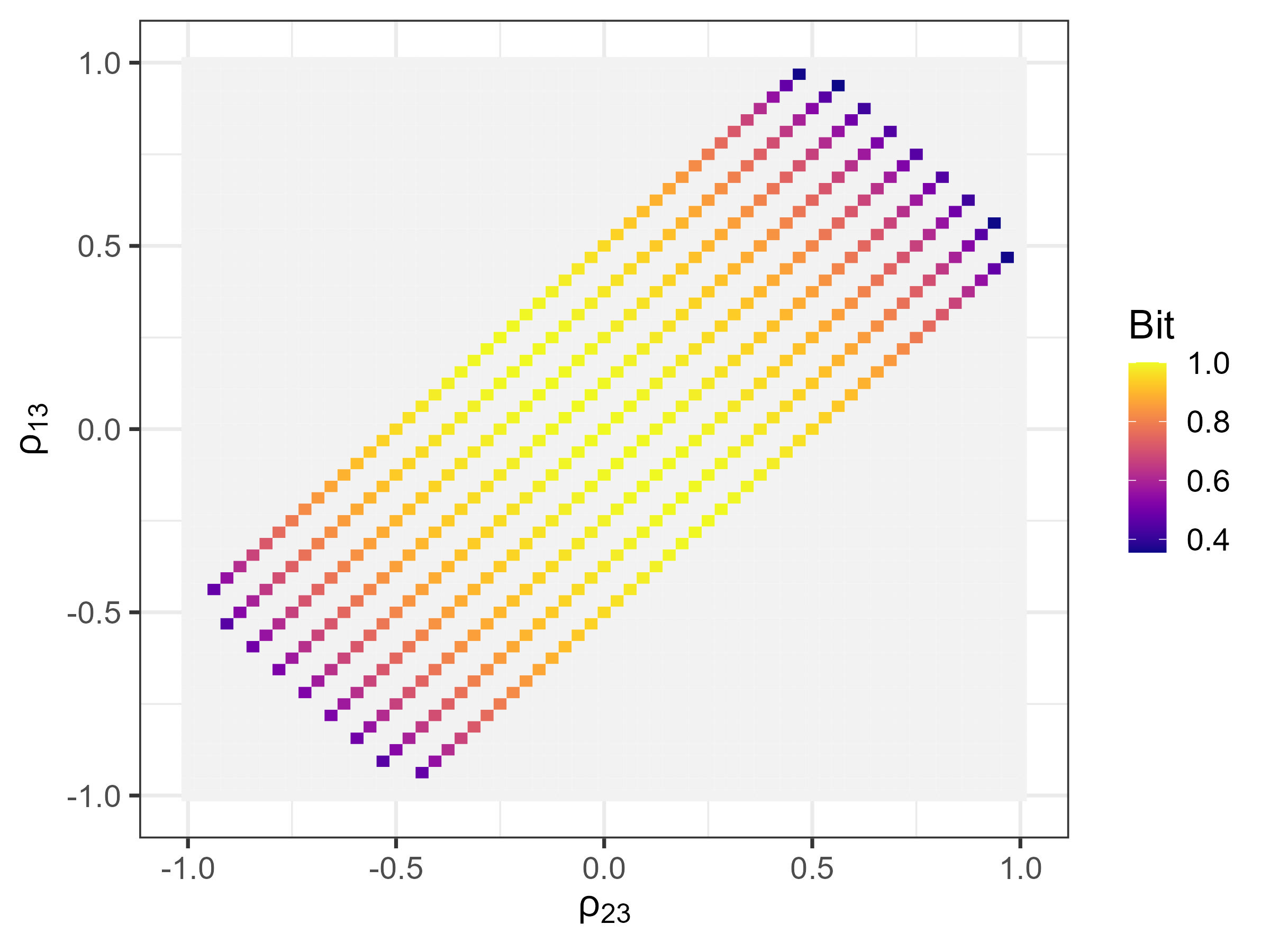}
  \caption{S-Entropy : $r_1=r_2$ and $\rho_{12}=0.5$}
  \label{fig:4}
\end{subfigure}
\caption{Min-entropy and Shannon entropy when knowing 2 challenges}
\label{fig:entropy_know2}
\end{figure*}

Figure \ref{fig:entropy_know2} shows the Min and Shannon entropy for different $\rho_{ij}$ values for an arbiter PUF with $n=32$. Again, we see that the remaining APUF entropies depend on the correlations between the two known challenges and the other challenges. Once again, the min entropy is more dramatically reduced than the Shannon entropy, as expected. This means that there is an increase in the probability of guessing a third challenge correctly when two challenges are known. On the diagonal (yellow) the two known challenges do not reveal any information about the remaining challenges. While one CRP might give us some information about the unknown challenge, the other CRP leads us on the opposite direction, so we do not learn anything and the probability to guess the unknown challenge response is still $0.5$, i.e. has entropy 1 bit.

\subsection{Neighborhoods}
We now ask how we may generalize the notion of similarity bins introduced earlier, to predictability and entropy neighborhoods. 
The question is which challenges ${\bf \Phi_3}$ lie within a certain "distance" / ``semi-metric'' $d$ to two anchor challenges rather than one. We would like an algorithm to enumerate the challenges that lie within this predictability or entropy neighborhood, denoted by $B(d, {\bf \Phi_1, \Phi_2}, r_1,r_2)$, and as a by-product, its size.  To this end, define

\begin{align}
&B(A, {\bf \Phi_1, \Phi_2}, r_1, r_2) = \bigl\{ {\bf \Phi_3}: \label{eq:Bk2pred1}\\
& \;\;\;  \max(  P(R_{\bf \Phi_3}= 1|R_{\bf \Phi_1}=r_1, R_{\bf \Phi_2}=r_2),\nonumber \\
& \;\;\; 1-P(R_{\bf \Phi_3}= 1|R_{\bf \Phi_1}=r_1, R_{\bf \Phi_2}=r_2)) =A \bigr\}\nonumber  \\
&B(H, {\bf \Phi_1, \Phi_2}, r_1, r_2) = \bigl\{ {\bf \Phi_3}:\\
& \;\;\;  H(R_{\bf \Phi_3}|R_{\bf \Phi_1}=r_{\bf \Phi_1}, R_{\bf \Phi_2} =r_{\bf \Phi_2}) =H \bigr\}
  \nonumber  
\end{align}

In this ``know 2, predict a third'' case, we will see that not all prediction accuracies/entropies are possible. It depends on the relationship between the challenges you know. If you know two challenges, they are fixed, and hence so is their correlation $\rho_{12}$. We are thus interested in seeing how many challenges lie at different correlations $\rho_{13}, \rho_{23}$ to these challenges.

In order to define the similarity neighborhoods and count them, we will need to quantify which correlation triples $(\rho_{12}, \rho_{13}, \rho_{23})$ are possible. 

\begin{figure}[h!]
		\centering
		\includegraphics[width=0.75\linewidth]{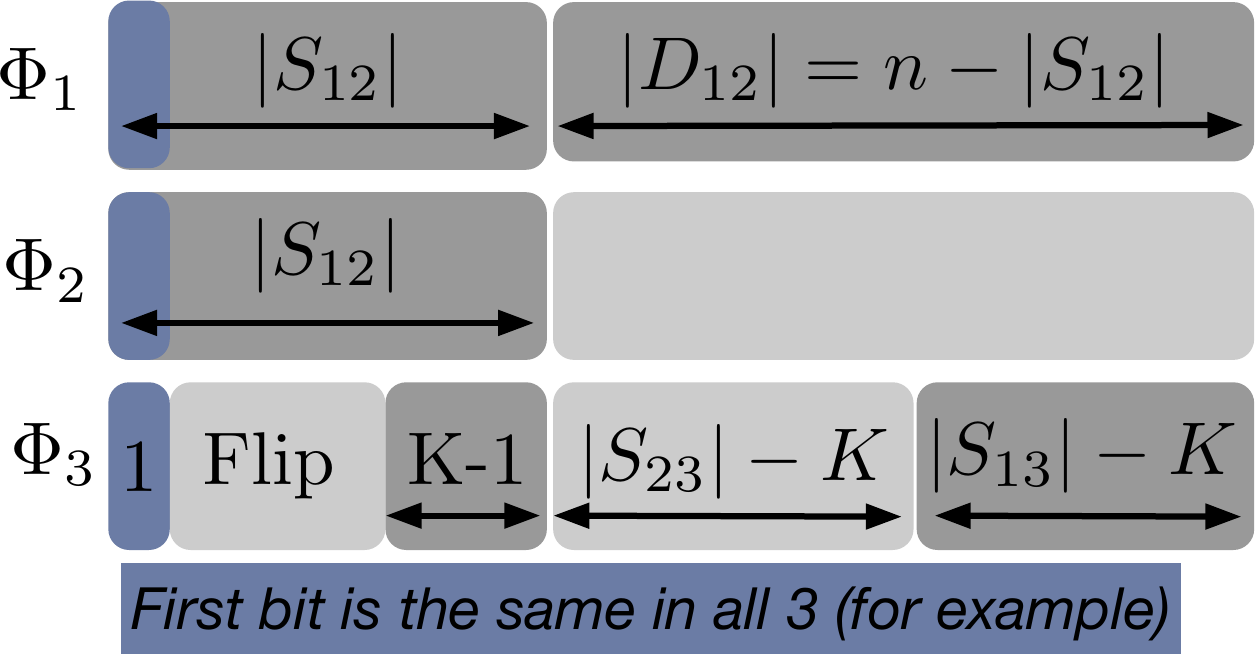}
		\caption{Illustration of the sizes of the difference index sets $S_{12}, S_{13}, D_{12}, S_{23}$ and integer $K$ representing the number of indices in $S_{12}$ that we keep the same in ${\bf \Phi_3}$. In this example we assume the first bit is the same in all three ${\bf \Phi_{1,2,3}}$.}% (this need not be the case, just for illustration).}
		\label{fig:illu_gb2anch}
\end{figure}

This is illustrated in Figure \ref{fig:illu_gb2anch}. To approach this,  define 
\begin{align}
    K &: = \text{ number of indices in } S_{12}\text{ that we ``keep''  in }{\bf \Phi_3} \nonumber \\
    & \overset{(a)}{=} \frac{|S_{13}|+|S_{23}|+|S_{12}|-n}{2} \label{eq:keep}
\end{align}

where (a) follows by setting $x: = |S_{13}|-K$ (the number of bits from $D_{12}$ that belongs to $|S_{13}|$), and $y: = |S_{23}|-K$ (the number of bits from $D_{12}$ that belongs to $|S_{23}|$), from which, since $x$ and $y$ must cover the whole $D_{12}$ set we see that $x+y=|D_{12}|=n-|S_{12}|$ and by replacing $x$ and $y$, we obtain $K=\frac{|S_{13}|+|S_{23}|+|S_{12}|-n}{2}$.

\subsubsection{Neighborhood construction} How can we find one challenge in $B(d, {\bf \Phi_1, \Phi_2}, r_1, r_2)$?
We enumerate the steps below which takes known challenges ${\bf \Phi_1, \Phi_2}$ with correlation coefficient $\rho_{12}$ and produces a single output challenge ${\bf \Phi_3}$ at the desired distance $d$ in a chosen ``semi-metric'' space (accuracy, Shannon or min entropies).

This distance $d$ depends on the challenge response values $r_1$ and $r_2$, and how to pick $\rho_{13}, \rho_{23}$ to satisfy this equation:
\begin{enumerate}
    \item Find $\rho_{13}, \rho_{23}$ based on desired distance $d$ and the set of all possible $(\rho_{12}, \rho_{13}, \rho_{23})$ tuples, given by Figure for the given $\rho_{12}$. How this Figure is obtained is presented in the Question below ``Which correlations $\rho_{13}$ and $\rho_{23}$ are possible given $\rho_{12}$''. This essentially boils down to picking $\rho_{13}, \rho_{23}$ of the desired accuracy (represented by color) in our numerical evaluations. 
    \item The $\rho_{12}, \rho_{13}, \rho_{23}$  determine the sizes $|S_{12}|, |S_{13}|, |S_{23}|$ of the sets $S_{12}, S_{13}, S_{23}$. From here we can use Algorithm \ref{alg:2anch_CC} in Appendix \ref{app:bin} which gives the following steps :
    \item From $|S_{12}|, |S_{13}|, |S_{23}|$  we find $K$ as in \eqref{eq:keep}. Select any $K$ indices in $S_{12}$ to keep fixed in ${\bf \Phi_3}$. The remaining indices in $S_{12}$ in ${\bf \Phi_3}$ must be flipped. 
    \item For the remaining indices in $D_{12}$, pick $|S_{13}| - K$ of them the same as ${\bf \Phi_1}$ and flip the remaining ones with respect to ${\bf \Phi_1}$, yielding the desired $\rho_{13}$ and $\rho_{23}$.
\end{enumerate}

\subsubsection{Example: constructing a ${\bf \Phi_3} \in B(d, {\bf \Phi_1, \Phi_2}, r_1, r_2)$} 
 Given challenges ${\bf \Phi_1}$ and ${\bf \Phi_2}$, to create a third vector with desired correlations (note that not all will be possible), the main idea is to know how many bits we have to keep/fix from the first two challenges and how many we have to flip. 
Consider $n=8$ and ${\bf \Phi_1}:=(+1,+1,+1,+1,+1,+1,+1,+1,+1)$, ${\bf \Phi_2}:=(+1,+1,+1,+1,-1,-1,-1,-1,+1)$, then $\rho_{12}=0$. Say we wish to predict a third challenge ${\bf \Phi_3}$ with probability of accuracy $A\approx0.6$.  We follow the steps of the algorithm:

\begin{enumerate}
    \item By generating the Figures from \eqref{eq:rhopossible1} -- \eqref{eq:rhopossible4}, we can see that $\rho_{23}=0$, $\rho_{23}=1/3$ and $\Phi_{1,1}=\Phi_{2,1}=\Phi_{3,1}=1$ is one of the multiple possible choices. %satisfying this accuracy.
    \item This is equivalent to selecting  $|S_{23}|=4$ and $|S_{13}|=6$.
    \item To create a challenge with these desired correlations to the two known challenges, we need to decide how many indices in ${\bf \Phi_3}$ to keep from $S_{12}$, let us call this set of indices $K$, and then how many to keep and flip from the set $D_{12}$.
Since $\Phi_{1,1}=\Phi_{2,1}=\Phi_{3,1}=1$ the first bit belongs to the sets $S_{12},\,S_{13},\, S_{23}$ so we have to select it in the ones we fix and $K-1=(|S_{12}|+|S_{13}|+|S_{23}|-n)/2-1=(4+4+6-8)/2-1=2$ other bits to fix from the $S_{12}$. Take for example ${\bf \Phi_3} :=(+1,*,+1,+1,*,*,*,*,+1)$ where the $*$ positions still need to be filled in.
    \item Finally, we can flip the other bits in $S_{12}$  to obtain ${\bf \Phi_3}:=(+1,-1,+1,+1,*,*,*,*,+1)$. Then we select $|S_{13}|-K=3$ bits from $D_{12}$ to fix and flip the left ones as  ${\bf \Phi_3}:=(+1,-1,+1,+1,+1,+1,+1,-1,+1)$. Equivalently, we could have fixed $|S_{23}|-K=1$ bits from $D_{12}$ and flipped  the left ones. Then, this ${\bf \Phi_3}$ can be predicted with accuracy $P(R_{\bf \Phi_3}= 1|R_{\bf \Phi_1}=1, R_{\bf \Phi_2}=1) =  \frac{1}{2}\left[1+ \frac{\arcsin{\frac{1}{3}} + \arcsin{0} }{\frac{\pi}{2}+\arcsin{0}}\right]\approx 0.61$. 
\end{enumerate}	

The algorithm is given in Algorithm \ref{alg:2anch_CC} in Appendix \ref{app:bin}.

\subsubsection{Neighborhood size} 
So far we have produced one challenge of a given desired distance to the anchors. We now answer how many challenges exist with that ditance, and how can they be efficiently enumerated? The answer to this is derived directly from the way we created the single challenge -- i.e. by looking at how many arbitrary choices we had. For example, in the case $\phi_{1,1}=\phi_{2,1}=\phi_{3,1}$: how many ways are there to choose $K-1$ bits from $S_{12}-1$ (since the first bit is fixed in this example)  and $|S_{13}|-K$ bits from $D_{12}$, with $|D_{12}|=n-|S_{12}|$. Now, there are two possibilities, when $\phi_{1,1}=\phi_{2,1}$ and when $\phi_{1,1}\neq \phi_{2,1}$ which must be treated differently as the 1st position is  special in the $\Phi$ notation. %So, we break apart the counting into the following cases:
\begin{itemize}[leftmargin=*] 
    \item If $\phi_{1,1}=\phi_{2,1}$
and $|S_{13}|+|S_{23}|+|S_{12}|-n\equiv 0 \mod 2$, then do not flip first bit: 
        \begin{align}\label{eq:density1}
  \#\text{of challenges }&=  {|S_{12}|-1 \choose K-1}{n-|S_{12}| \choose |S_{13}|-K}\\
  &={|S_{12}|-1 \choose K-1}{n-|S_{12}| \choose |S_{23}|-K}
\end{align}
\item If $\phi_{1,1}=\phi_{2,1}$ and $|S_{13}|+|S_{23}|+|S_{12}|-n\equiv 1 \mod 2$, then flip the first bit: 
\begin{align}\label{eq:density2}
  \#\text{of challenges }&=  {|S_{12}|-1 \choose K}{n-|S_{12}| \choose |S_{13}|-K}\\
  &={|S_{12}|-1 \choose K}{n-|S_{12}| \choose |S_{23}|-K}
\end{align}
    \item If $\phi_{1,1}\neq\phi_{2,1}$ we need $|S_{13}|+|S_{23}|+|S_{12}|-n\equiv 0 \mod 2$ since we necessarily have $\phi_{3,1}\neq\phi_{2,1}$ or $\phi_{3,1}\neq\phi_{1,1}$
    \begin{align} \label{eq:density3}
  &\#\text{of challenges }\\
  &=  {|S_{12}| \choose K}{n-|S_{12}| \choose |S_{13}|-K-1}\text{ if ${S}({\bf \Phi_1, \Phi_3})\equiv 0 \mod 1$}\\
  &={|S_{12}| \choose K}{n-|S_{12}| \choose |S_{23}|-K-1} \text{ if ${S}({\bf \Phi_2, \Phi_3})\equiv 0 \mod 1$} 
\end{align}
\end{itemize}	

Notice that, contrary to the accuracy, the number of challenges does not depend on the challenge responses.
To create  similarity bins of size $m$ with two anchors we can use Algorithm \ref{alg:2anch_main}, which in turn calls Algorithms \ref{alg:2anch_CBSS} and \ref{alg:2anch_CBDS}.

\subsubsection{Neighborhood landscape discussion}
{Now we provide some in-depth analysis on the details of a neighborhood's landscape with the following questions.}
\paragraph{Which correlations $\rho_{13}$ and $\rho_{23}$ are possible given $\rho_{12}$?} The answer depends on whether $\phi_{1,1}, \phi_{2,1}, \phi_{3,1}$ are the same or different. This is because this first bit has a different variance than the others, as per Lemma \ref{lemma:w}. We present one case as an example; the others follow similarly.

Assume $\phi_1^1=\phi_2^1=\phi_3^1$. Then  equation \eqref{eq:keep} and \eqref{eq:density1} imply two sufficient and necessary conditions on the correlation between the three challenges to be able to create the third one.
In particular, by requiring $K\in \mathbb{N}$ in \eqref{eq:keep}, and $\#\text{of challenges }=  {|S_{12}|-1 \choose K-1}{n-|S_{12}| \choose |S_{13}|-K}={|S_{12}|-1 \choose K-1}{n-|S_{12}| \choose |S_{23}|-K}>0$ we obtain the following conditions:
\begin{align*}
    K \in \mathbb{N} \rightarrow |S_{13}|+|S_{23}|+|S_{12}|-n & \equiv 0 \mod 2\\
 {|S_{12}|-1 \choose K-1}{n-|S_{12}| \choose |S_{13}|-K}>0 &\rightarrow  |S_{12}|\geq K\\
    & \rightarrow n-|S_{12}|\geq |S_{13}|-K\\
  { |S_{12}|-1 \choose K-1}{n-|S_{12}| \choose |S_{23}|-K} >0 & \rightarrow  n-|S_{12}|\geq |S_{23}|-K
\end{align*}
Re-writing these using correlation coefficients (recall $|S|=(\rho+1)\frac{n}{2}$) yields the inequalities relating the possible correlations $\rho_{12}, \rho_{13}, \rho_{23}$ (which also must all lie in $[-1,1]$):
\begin{align}
 %   \rho=\frac{2|S|}{n}-1&\leftrightarrow|S|=(\rho+1)\frac{n}{2}\\
    &\rho_{13}\leq -\rho_{23}+1-\rho_{12} \label{eq:rhopossible1}\\
    &\rho_{13}\geq \rho_{23}-1+\rho_{12}
    \end{align}
\begin{align}
    &\rho_{13}\leq \rho_{23}+1-\rho_{12}\\
    &\rho_{13}\geq -\rho_{23}-1+\rho_{12} \label{eq:rhopossible4}
\end{align}

\begin{figure*}\captionsetup[subfigure]{font=scriptsize}
\centering
\begin{subfigure}{0.25\textwidth}
  \includegraphics[width=\linewidth]{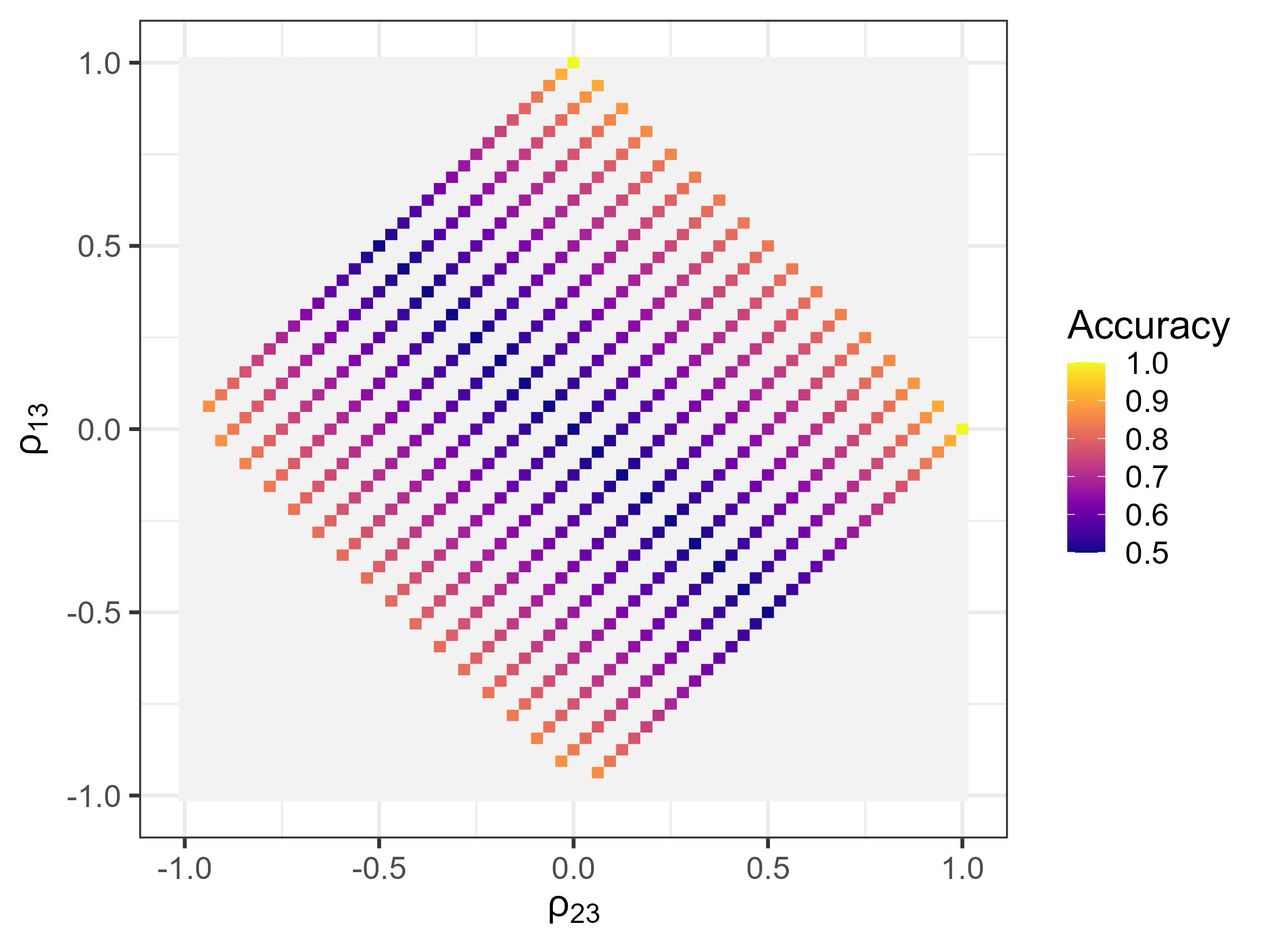}
  \caption{Accuracy : $r_1=r_2$ and $\rho_{12}=0$}
  \label{fig:1}
\end{subfigure}\hfil % <-- added
\begin{subfigure}{0.25\textwidth}
  \includegraphics[width=\linewidth]{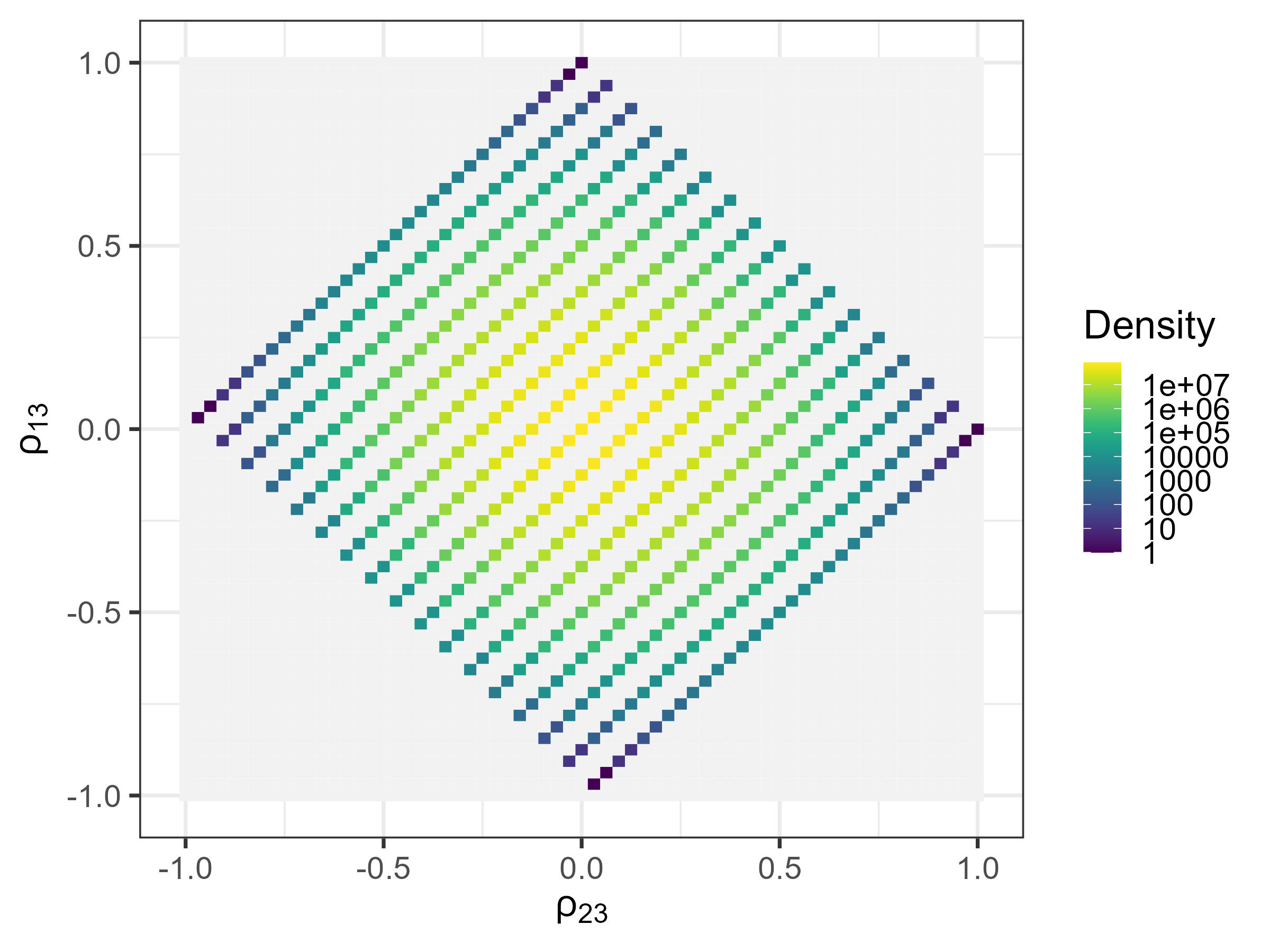}
  \caption{Density : $r_1=r_2$ and $\rho_{12}=0$}
  \label{fig:2}
\end{subfigure}\hfil % <-- added
\begin{subfigure}{0.25\textwidth}
  \includegraphics[width=\linewidth]{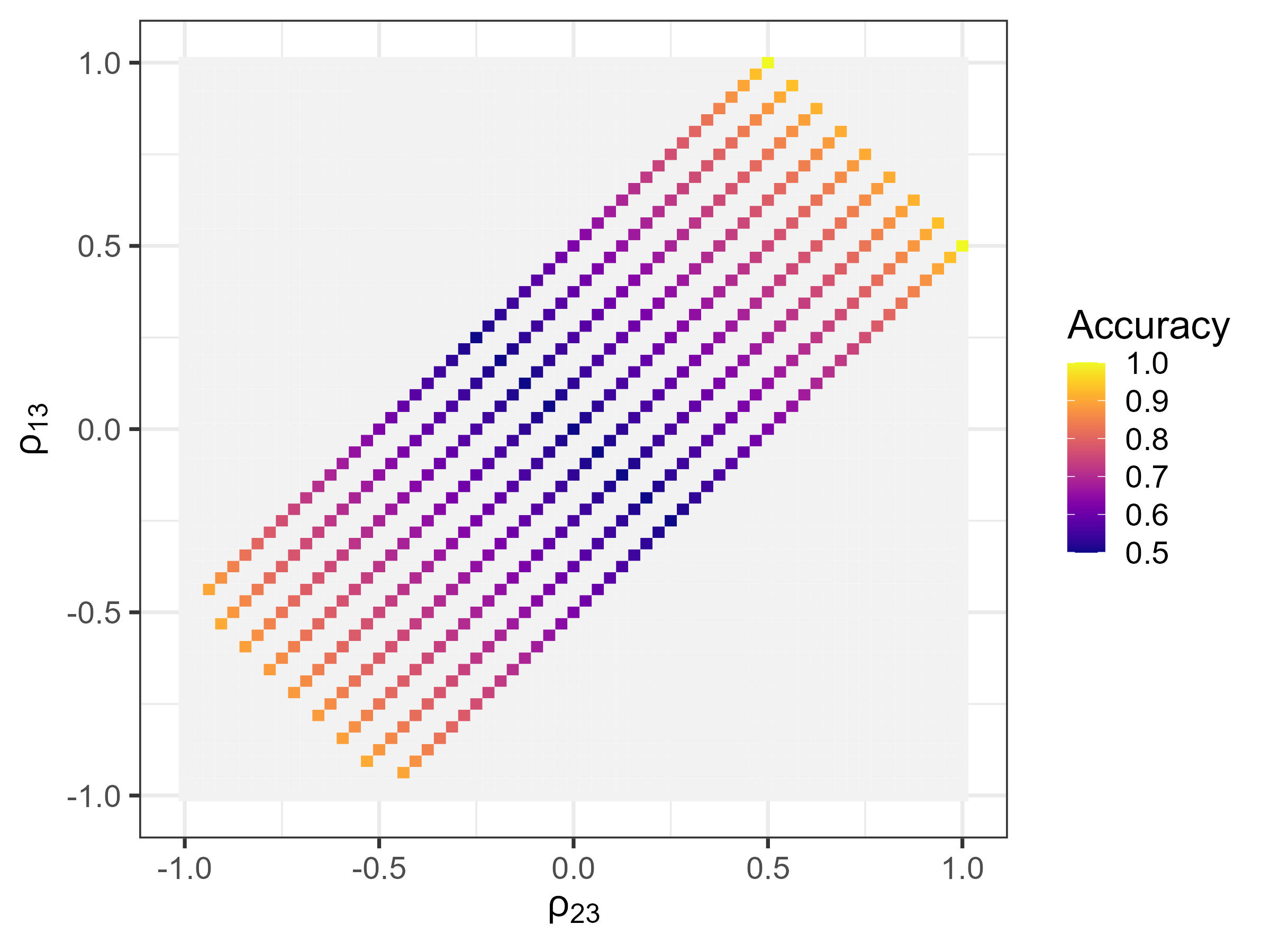}
  \caption{Accuracy : $r_1=r_2$ and $\rho_{12}=0.5$}
  \label{fig:3}
\end{subfigure}\hfil
\begin{subfigure}{0.25\textwidth}
  \includegraphics[width=\linewidth]{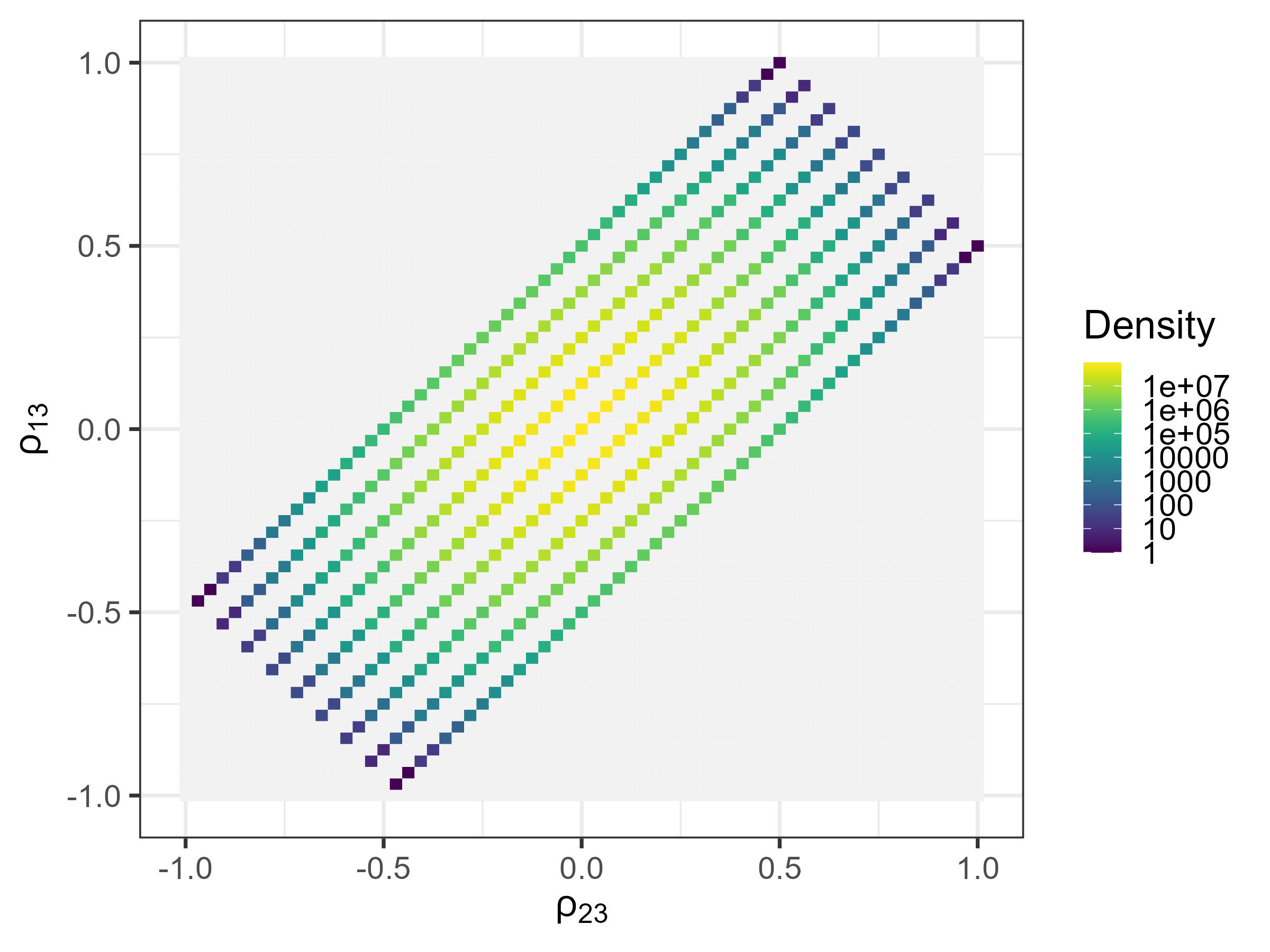}
  \caption{Density : $r_1=r_2$ and $\rho_{12}=0.5$}
  \label{fig:4}
\end{subfigure}

\medskip

\begin{subfigure}{0.25\textwidth}
  \includegraphics[width=\linewidth]{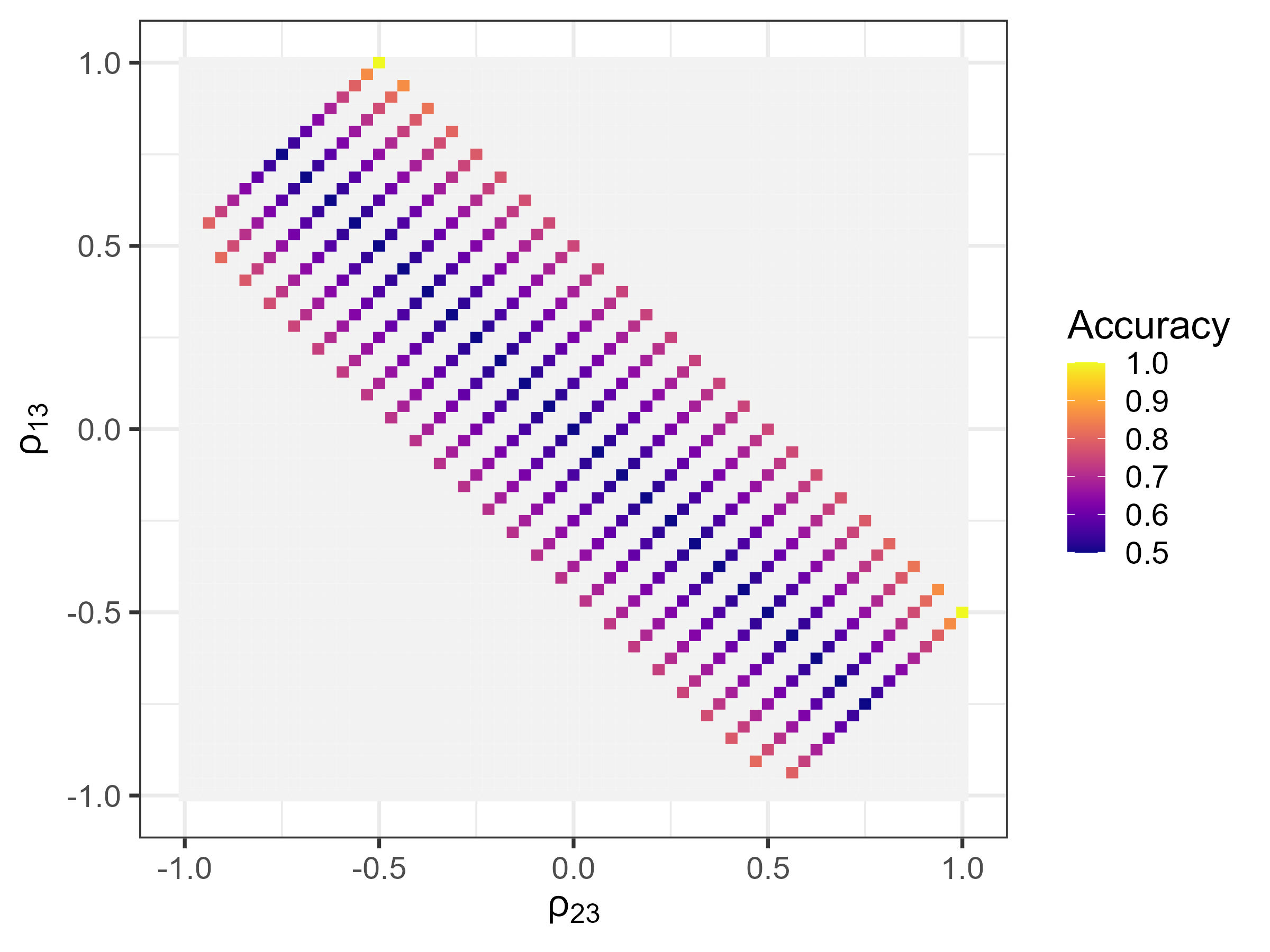}
  \caption{Accuracy : $r_1=r_2$ and $\rho_{12}=-0.5$}
  \label{fig:1}
\end{subfigure}\hfil % <-- added
\begin{subfigure}{0.25\textwidth}
  \includegraphics[width=\linewidth]{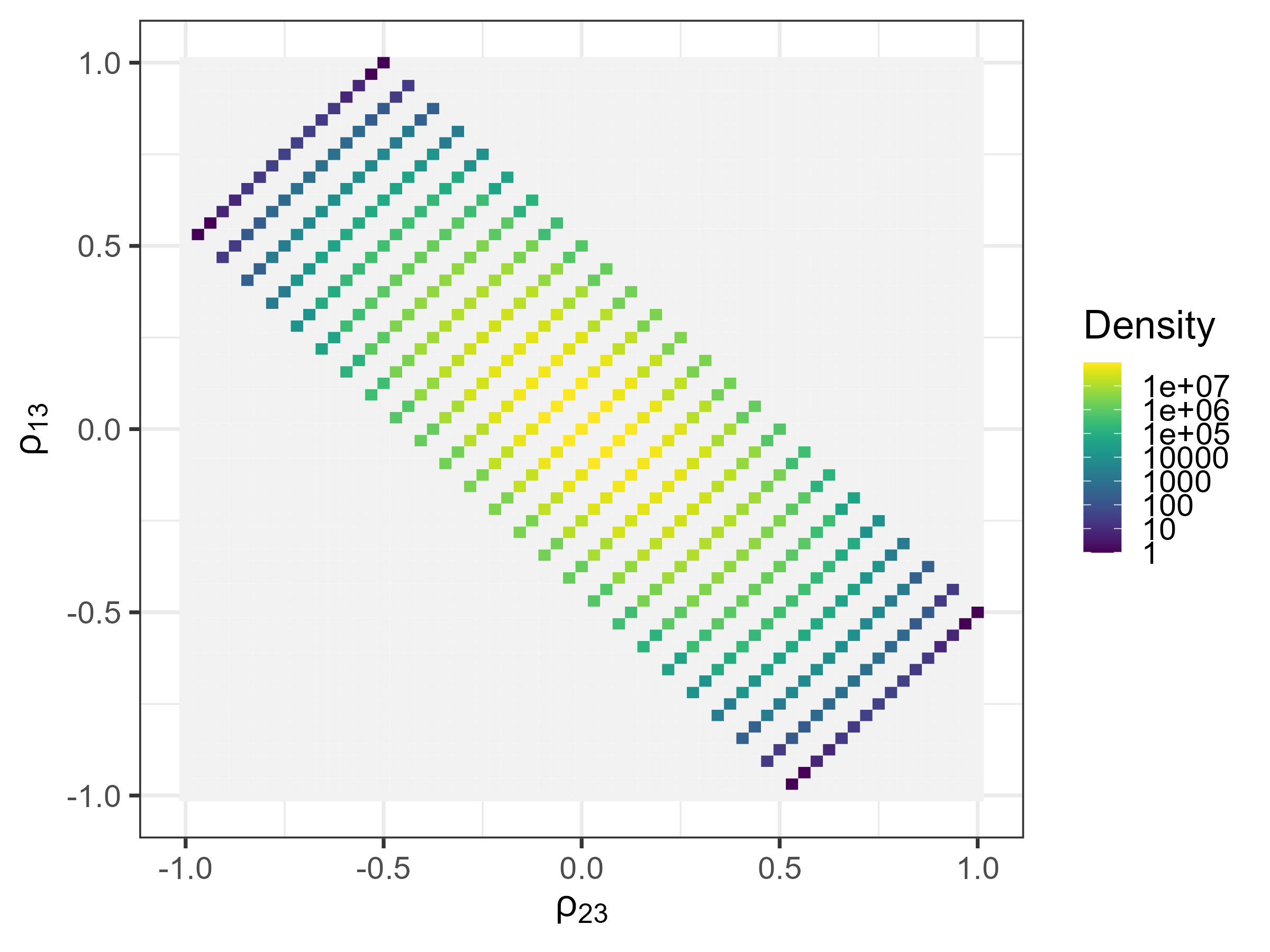}
  \caption{Density : $r_1=r_2$ and $\rho_{12}=-0.5$}
  \label{fig:2}
\end{subfigure}\hfil % <-- added
\begin{subfigure}{0.25\textwidth}
  \includegraphics[width=\linewidth]{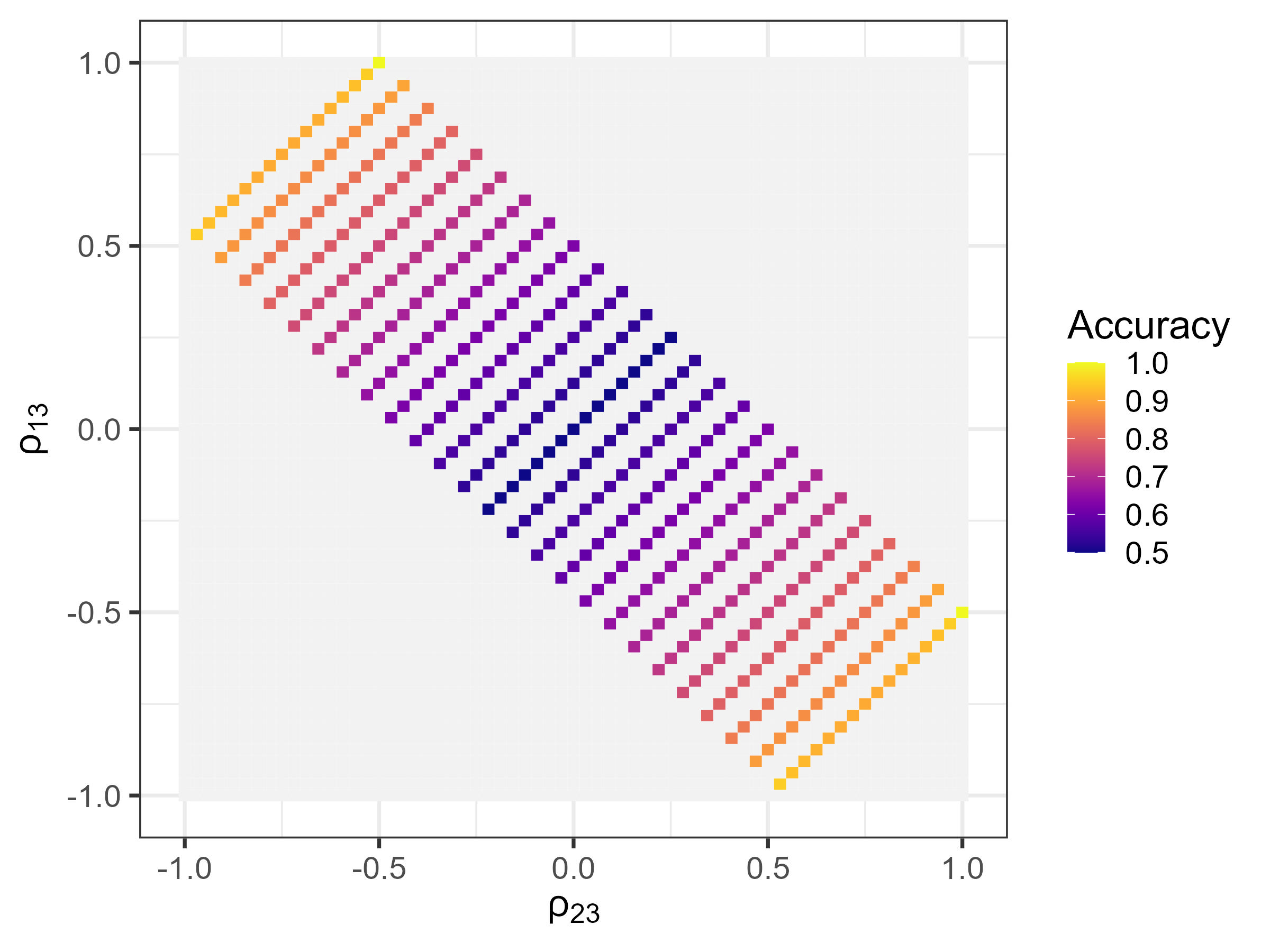}
  \caption{Accuracy : $r_1\neq r_2$ and $\rho_{12}=-0.5$}
  \label{fig:3}
\end{subfigure}\hfil
\begin{subfigure}{0.25\textwidth}
  \includegraphics[width=\linewidth]{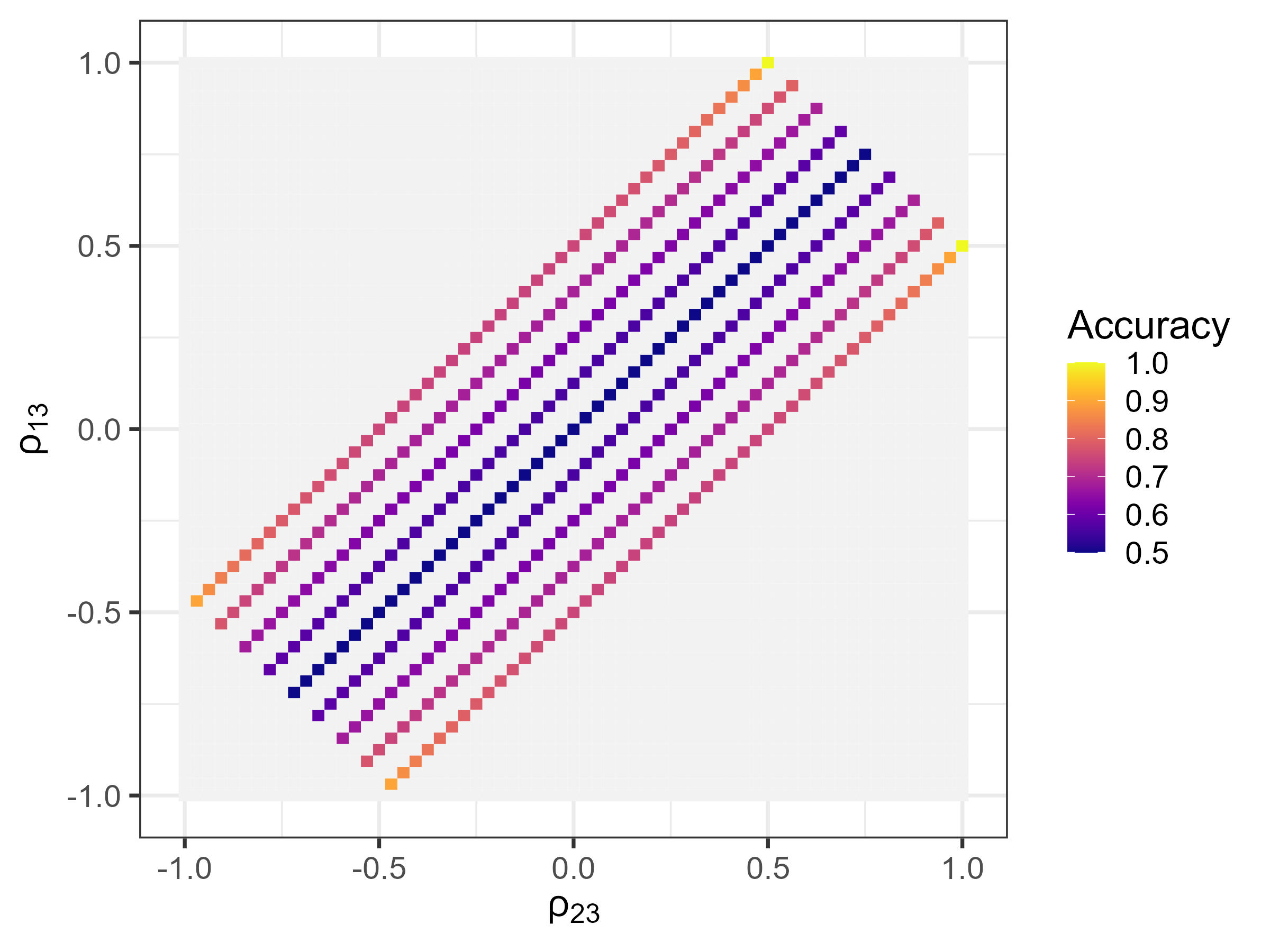}
  \caption{Accuracy : $r_1\neq r_2$ and $\rho_{12}=0.5$}
  \label{fig:4}
\end{subfigure}
\caption{Accuracy and density plot same and different responses. }
%\nrd{2 plots have been removed to fit in 14 pages}}
\label{fig:all_fig}
\end{figure*}

One can visualize these equations easily, yielding rotated rectangles in the $\rho_{13}, \rho_{23}$ plane for each given $\rho_{12}$ (fixed by the two known challenges ${\bf \Phi_1}$ and ${\bf \Phi_2}$. The shapes in Figures  \ref{fig:all_fig} indicate which triples are possible, as shown by the linear equations above. The largest range of possible correlations occurs when $\rho_{12}=0$, i.e. the first two challenges are uncorrelated. There are many such challenges, exactly how many is given by for example Figure \ref{fig:all_fig}. 
\paragraph{How many challenges lie at different possible $(\rho_{13}, \rho_{23})$ from given challenges with correlation $\rho_{12}$?} To obtain this, we simply evaluate Equations \eqref{eq:density1} -- \eqref{eq:density3} depending on what case we are in for each possible $(\rho_{12}, \rho_{13}, \rho_{23})$ triple. These yield the different colors in the yellow/blue plots of Figures \ref{fig:all_fig}. Notice that the densities do not depend on the actual values $r_1, r_2$ that the challenges take on.

\paragraph{What prediction accuracies and entropies are possible with different $(\rho_{12}, \rho_{13}, \rho_{23})$ triples?} Finally, given that we now know how many challenges there are at different correlations to one another, the question is how many challenges there are at different prediction accuracies or entropies. This is given by Theorem \ref{thm:k2p1} and shown in the purplish plots of Figures \ref{fig:entropy_know2} and \ref{fig:all_fig}.

\paragraph{Numerical evaluations and interpretations of plots} 
Algorithms \ref{alg:2anch_main}--\ref{alg:2anch_CBDS} in Appendix \ref{app:bin} allow us to create the bins $B(d, {\bf \Phi_1, \Phi_2}, r_1,r_2)$ for all possible distance $d$ in the ``semimetric'' space defined by the accuracy or the entropies, and then the  Figures \ref{fig:entropy_know2} and \ref{fig:all_fig}. We now present some numerical evaluations to provide an understanding of the similarity bins when we know 2 challenges and wish to understand how many challenges lie at different distances to these two. To do so, we illustrate the following three questions:

\begin{enumerate}
    \item Which correlations $\rho_{13}$ and $\rho_{23}$ are possible given $\rho_{12}$? This is given by the shape in Figures \ref{fig:all_fig}--\ref{fig:all_fig2} which depict equations \eqref{eq:rhopossible1} -- \eqref{eq:rhopossible4}.    
    \item What prediction distances are possible with different $(\rho_{12}, \rho_{13}, \rho_{23})$ triples? This is given by Theorem \ref{thm:k2p1} and is given by the  purplish plots of Figures \ref{fig:entropy_know2},\ref{fig:all_fig} and \ref{fig:all_fig2}.. The accuracy and entropies do depend on the actual values of the responses as well as the correlations. This may be intuitively thought of as follows: if the two challenges we know are well  ``aligned'', i.e. when $r_1\cdot r_2\cdot \text{sign}(\rho_{12})=+1$ this means the challenges have the same response and are highly correlated then they act more like one challenge with respect to the third one. Otherwise the information provided by the two challenges when  $\rho_{12}$ negative, $r_1=r_2$ is somewhat contradictory, then information is neutralized for $\rho_{13}=\rho_{23}$. The ``blue line'' on the accuracy graphs shows where the information is neutralised: line $y=-x$ if $r_1=r_2$, $y=x$ if $r_1\neq r_2$. 
    \item  How many challenges lie at different possible $(\rho_{13}, \rho_{23})$ from given challenges with correlation $\rho_{12}$?  This is given by equations \eqref{eq:density1} -- \eqref{eq:density3} and is given by the blue/yellow plots of Figures \ref{fig:all_fig}.
\end{enumerate}

\paragraph{Expected conditional entropy and accuracy of a challenge's response} 
Knowing the values of the entropies and accuracy in each neighborhood as well as the number of challenges in all of them. we can compute the expected conditional entropy and accuracy of a challenge's response.
Calling $d_i$ the distance in $B(d_i, {\bf \Phi_1, \Phi_2}, r_1,r_2)$ in the entropy or accuracy space and $n^2$ the maximum number of different neighborhoods knowing that if one is not possible then $|B(d_i, {\bf \Phi_1, \Phi_2}, r_1,r_2)|=0$.  The expected semi-metric $\bar{d}$ is then given as follows:
\begin{equation}
    \Bar{d}=\frac{1}{\#\text{possible challenges}}\sum_{i=1}^{n^2} d_i|B(d_i, {\bf \Phi_1, \Phi_2}, r_1,r_2)|.
\end{equation}

Table \ref{tab:expect_entropy_know2} shows the different expected conditional entropy and accuracy of a challenge's response in function of the correlation $\rho_{12}$ between the two anchors and the number of stages of the PUF, $n$.

There is quite a bit of information packed into these Figures. Some things to note include: highly predictable challenges have the yellow color in the left (a) plots: they tend to be around the edges and there are relatively few of them. Less predictable challenges are found along the dark blue/purplish lines in the left (a) plots and there tend to be many of them. However, sometimes we can see large clumps (density plots are greenish) that are also reasonably well predicted (pinkish). Overall, the hope is that such plots will be useful to PUF protocol design engineers to give them an idea of the landscape of challenges: how many there are at different prediction accuracies or conditional response entropies once one or two challenges are exposed. 

% \begin{figure}[h!]
% \centering
% \includegraphics[width=1\linewidth]{figures/all_in_one2.pdf}
% \caption{$r_1\cdot r_2\cdot \text{sign}(\rho_{12})=+1\text{ (a); }-1\text{ (b)}$.}
% \label{fig:all_fig2}
% \end{figure}

\begin{table*}[t]  
        \begin{tabular}{lrrrrrrrrr}\toprule
            &\multicolumn{3}{c}{\textbf{n=32}}&\multicolumn{3}{c}{\textbf{n=64}}&\multicolumn{3}{c}{\textbf{n=128}}
            \\\cmidrule(r){2-4}\cmidrule(r){5-7}\cmidrule(r){8-10}   
            &$\rho_{12}=-0.5$&$\rho_{12}=0$&$\rho_{12}=0.5$&$\rho_{12}=-0.5$&$\rho_{12}=0$&$\rho_{12}=0.5$&$\rho_{12}=-0.5$&$\rho_{12}=0$&$\rho_{12}=0.5$\\\midrule
            H-min    & 0.8259 & 0.8329 & 0.8450
                    & 0.8728 & 0.8788 & 0.8879
                    & 0.9080 & 0.9127 & 0.9194\\
            H-Shannon   & 0.9798 &0.9815 & 0.9842
                    & 0.9898 & 0.9908 & 0.9922
                    & 0.9948 & 0.9954 & 0.9961\\
            Accuracy& 0.5663 &0.5634 & 0.5584 
                    & 0.5472 & 0.5448 & 0.5413
                    & 0.5335 & 0.5317 & 0.5292
            \\\bottomrule
        \end{tabular}
        \caption{Expected conditional entropies and accuracies of a challenge's response knowing two CRPs}\label{tab:expect_entropy_know2}
\end{table*}

\section{Conclusion}

\noindent We have presented two new tools for understanding and exploiting the CRP correlations of APUF (and by almost immediate extension, compute then entropy). The first is the response similarity, or the probability that two challenges yield the same response, which is a function of the challenges themselves. This may be used to optimally predict the response to one challenge given the response to another challenge, and to obtain the conditional response entropy, The second is the derivation of a simple algorithm for enumerating the similarity bins, or entropy bins to a given challenge, i.e. all challenges that have the same response similarity or conditional response entropy to a given anchor challenge. 
In combination, these form a powerful tool for understanding the CRP landscape: how CRPs are correlated and may be used to obtain statistical properties of linear-threshold-based PUFs.

\section*{Declarations}
\begin{itemize}[leftmargin=*]
	\item	{\it Ethical Approval:} This work is original and has not been published elsewhere, nor is it currently under consideration for publication elsewhere.
    \item	{\it Competing interests:} The authors have no relevant financial or non-financial interests to disclose.
    \item	{\it Authors' contributions:} All authors contributed to the study conception and design. Material preparation, simulations and analysis were performed by Vincent Dumoulin, Natasha Devroye and Wenjing Rao. The first draft of the manuscript was written by Vincent Dumoulin and Natasha Devroye then all authors commented on previous versions of the manuscript. All authors read and approved the final manuscript.
    \item	{\it Funding:} This work was supported by the National Science Foundation under award 2217023 and 2244479.
    \item	{\it Data:} No exterior data set is used. All simulations and plots are derived based on the statistical models, equations and pseudo algorithms presented in the paper.
 \end{itemize}

%%%% 8. BILBIOGRAPHY %%%%
\bibliographystyle{IEEEtran}
\bibliography{main}

\newpage

\begin{appendix}
\section*{Algorithms for bin creation}\label{app:bin}
Algorithm \ref{alg:bin} builds a similarity bin. For a different ``semimetric'', use the theoretical results or the graph in Figure \ref{fig:entropy_know1} to get the two ${\bf s}=(s_1,s_2)$ factors that will give you the desired value/ create a similarity bin for each $s_i$ (two similarity factors for entropy and accuracy). Form the union of the two disjoint bins.
\begin{algorithm}[h]
	 \caption{Bin construction for one anchor: \label{alg:bin}}
			\KwIn{anchor ${\bf \Phi}, n$, similarity factor(s) ${\bf s}=(s_1,s_2)$ derived from a wanted ``semimetric" (entropy, accuracy,probability).}
			\KwOut{Bin $B({\bf s}, {\bf \Phi})$.}
		
 \tcp{initialize}
 
			$B \leftarrow \emptyset$
			
			\tcp{use $s$ and $\phi_1$ to determine $\phi'_1$. }
			\eIf{$s$ mod $1 != 0$}{ 
				$\phi'_1 = -\phi_1$ \tcp*{$s$ even}}
 {	$\phi'_1 =\phi_1$ \tcp*{$s$ odd}}
			\tcc{use $s$ to decide how many of $\phi_2, \cdots, \phi_n$, to flip to construct $\phi'_2, \cdots, \phi'_n$. }
% 			$k \leftarrow \lfloor S/2 \rfloor$${\bf \Phi'}$
			
			$F \leftarrow \{f| f \in \{0, 1\}^{n-1}, \sum{f_i} = \lfloor s \rfloor \}$ 
			
			\tcc{ $F$ contains all the strings of size $n-1$ with Hamming weight $\lfloor s \rfloor$, and 
			$f$ encodes where to flip $\phi_{2}, \cdots, \phi_n$ for $\phi'_{i}$'s}
			\For{$f \in F$ 
			}{
			 \For{i = 2 \KwTo n}{

			 \eIf{$f_{i-1}==0$}
			 {$\phi'_{i} = \phi'_{i}$}
			 {$\phi'_{i} = - \phi'_{i}$}
			 }
			 ${\bf \Phi'} \leftarrow (\phi'_1, \phi'_2, \cdots, \phi'_n, +1)$

			 $B \leftarrow B \bigcup \{{\bf \Phi'}\}$
			 	}
			\KwRet{$B$} \tcp{This is $B({\bf s}, {\bf \Phi})$}
\end{algorithm}

\begin{algorithm}[h]
	\caption{Bin construction with two anchors\label{alg:2anch_main}}
	\KwIn{Anchors ${\bf \Phi_1, \Phi_2}$ and responses $r_1$ $r_2$}
	\KwOut{Similarity bins of size $m$ for all possible distance $d$ : $B(d, {\bf \Phi_1, \Phi_2}, r_1,r_2)$}

 \tcp{initialize: look for the bits to select}
 	\For{$|S_{12}| = 1$ \KwTo\, $n$ }{
 	    \For{$|S_{13}| = 1$ \KwTo\, $n$}{
 	        \For{$|S_{23}| = 1$ \KwTo\, $n$}{
 	        ${\bf K} \leftarrow (|S_{13}|+|S_{23}|+|S_{12}|-n)/2$\;
 	        \uIf{$\Phi_{1,1}=\Phi_{2,1}$}{
 	        CBSS$({\bf \Phi_1, \Phi_2},\, r_1,\, r_2,\, |S_{12}|,\,|S_{13}|,\,|S_{23}|)$
 	    }

 	         \uIf{$\Phi_{1,1}\neq\Phi_{2,1}$}{
 	         CBDS$({\bf \Phi_1, \Phi_2},\, r_1,\, r_2,\, |S_{12}|,\,|S_{13}|,\,|S_{23}|)$
 	    }	
        \lElse{Ignore \tcp{Challenge does not exist}\DontPrintSemicolon}}}
 \KwRet{$B(d, {\bf \Phi_1, \Phi_2}, r_1,r_2)$ for all distances $d$} 
}
\end{algorithm}

Algorithm \ref{alg:2anch_CC} builds a third challenge at a distance $d$ in a the chosen ``semimetric'' space (accuracy or entropy space). Algorithms \ref{alg:2anch_main}-\ref{alg:2anch_CBDS} create the bins using the previous algorithm.

\begin{algorithm}
	 \caption{CC: Create a 3rd Challenge ${\bf \Phi_3}$ \label{alg:2anch_CC}}
	\KwIn{Anchors ${\bf \Phi_1, \Phi_2}$, responses $r_1$ $r_2$, $|S_{12}|,\,|S_{13}|,\,|S_{23}|,\,\Phi_{3,1}$}
	\KwOut{a challenge ${\bf \Phi_3}$ with a desired distance d}

 \tcp{construct $\Phi_3$ from $\rho_{12},\,\rho_{13},\,\rho_{23}$ we know  $\Phi_{1,1}\stackrel{?}{=}\Phi_{2,1}\stackrel{?}{=}\Phi_{3,1}$}
 ${\bf K} \leftarrow (|S_{13}|+|S_{23}|+|S_{12}|-n)/2$\;
\uIf{$\Phi_{1,1}=\Phi_{2,1}=\Phi_{3,1}$}{
\eIf{${\bf K}\in \mathbb{N}$}{ 
			 Fix: the first bit and ${\bf K}-1$ other bits in $ S_{12}$\;
			 
			 Flip: all other bits in $S_{12}$\;
			 
			 Fix: any $|S_{13}|-K$ bits in $D_{12}$\; 
			 Flip: all other bits in $D_{12}$ \;
			  \tcp{equivalently fix $|S_{23}|-K$ bits in $D_{12}$, and flip the rest}}
            {\KwRet False\;  \tcp{\# of bits to fix is not an integer}}}
\uElseIf{$\Phi_{1,1}=\Phi_{2,1}\neq\Phi_{3,1}$}{
\eIf{${\bf K}=x.5$}{ 
			 Fix: any $\lfloor{\bf K}\rfloor$ bits except the 1st bit in $S_{12}$\;
			 
			 Flip: all other in $S_{12}$ (including the 1st bit)\;
			 
			 Fix: any $|S_{13}|-K$ bits in $D_{12}$\;
			 Flip: all other bits in $D_{12}$\;
			  \tcp{equivalently fix $|S_{23}|-K$ bits in $D_{12}$, and flip the rest}}
            {\KwRet False\;}}
\uElseIf{$\Phi_{1,1}\neq\Phi_{2,1}=\Phi_{3,1}$}{\eIf{${\bf K}\in \mathbb{N}$}{ 
			 Use: the first bit of $\Phi_{2,1}$\;
			 Fix: any ${\bf K}$ bits in $S_{12}$\;
			 
			 Flip: all other bits in $S_{12}$\;
			 
			Fix: any $|S_{23}|-K-1$ bits in $D_{12}$\;
% 			-\text{ first bit}$, fix those and 
            Flip: all other bits in $D_{12}$\;}
            {\KwRet False\;  \tcp{Number of bits to fix is not integer}}}
\uElseIf{$\Phi_{1,1}=\Phi_{3,1}\neq\Phi_{2,1}$}{\eIf{${\bf K}\in \mathbb{N}$}{ 
			 Use: the first bit of $\Phi_{3,1}$ \;
			 Fix: any ${\bf K}$ bits in $S_{12}$ \;
			 
			 Flip: all other bits in $S_{12}$ \;
			 
			 Fix: any $|S_{13}|-K-1$ bits in $D_{12}$ \;
			 %-\text{ first bit}$, fix those and 
			 Flip: all other bits in $D_{12}$ \;}
            {\KwRet False\;  \tcp{Number of bits to fix is not integer}}}
    \KwRet{$\Phi_3$} \tcp{Challenge $\Phi_3$ has desired distance}
\end{algorithm}

\begin{algorithm}[h]
	\caption{CBSS: Create Bin first bit Same Sign: $\Phi_{1,1}=\Phi_{2,1}$\label{alg:2anch_CBSS}}
	\KwIn{Anchors ${\bf \Phi_1, \Phi_2}$, responses $r_1$ $r_2$, $|S_{12}|,\,|S_{13}|,\,|S_{23}|$ under $\Phi_{1,1}=\Phi_{2,1}$}
	\KwOut{$B(d, {\bf \Phi_1, \Phi_2}, r_1,r_2)$ of size $m$}
	
	$\rho_{12}=(2|S_{12}|)/n-1$\;
 	        \Begin{$\Phi_{3,1}=\Phi_{1,1}$\;
 	            $\rho_{13}=(2|S_{13}|)/n-1$\;
 	            $\rho_{23}=(2|S_{23}|)/n-1$\;
 	            $\#\text{of challenges }=  {|S_{12}|-1 \choose K-1}{n-|S_{12}| \choose |S_{13}|-K}$\;
 	          %  ={S_{12}-1 \choose K-1}{n-S_{12} \choose S_{23}-K}$\;
 	            \uIf{${\bf K}\in \mathbb{N}\quad \&\quad \#\text{of challenges }\geq 0$}{ 
                $p\leftarrow \frac{1}{2}\left[1+\frac{r_1\arcsin{\rho_{13}}+r_2\arcsin{\rho_{23}}}{\pi/2+r_1r_2\arcsin{\rho_{12}}}\right]$\;
                $d_1=$ Accuracy(p) or entropy(p)\;
                \For{$i=1$ \KwTo\,$m$}{
                $\Phi_3 \leftarrow$ CC$({\bf \Phi_1, \Phi_2},\, r_1,\, r_2,\, |S_{12}|,\,|S_{13}|,\,|S_{23}|,\,\Phi_{3,1})$\;
                $B(d_1, {\bf \Phi_1, \Phi_2}, r_1,r_2)\leftarrow B(d_1, {\bf \Phi_1, \Phi_2}, r_1,r_2) \bigcup \{\Phi_3\}$}}
                \lElse{Ignore \tcp{Challenge does not exist}\DontPrintSemicolon}}
 	        \Begin{$\Phi_{3,1}=-\Phi_{1,1}$\;
 	            $\rho_{13}=(2|S_{13}|+1)/n-1$\;
 	            $\rho_{23}=(2|S_{23}|+1)/n-1$\;
 	            $\#\text{of challenges }=  {|S_{12}|-1 \choose K}{n-|S_{12}| \choose |S_{13}|-K}$\;
 	          %  ={|S_{12}|-1 \choose K}{n-|S_{12}| \choose |S_{23}|-K}$\;
 	            \uIf{${\bf K}\in \mathbb{N}\quad \&\quad \#\text{of challenges }\geq 0$}{ 
                $p\leftarrow \frac{1}{2}\left[1+\frac{r_1\arcsin{\rho_{13}}+r_2\arcsin{\rho_{23}}}{\pi/2+r_1r_2\arcsin{\rho_{12}}}\right]$\;
                $d_2=$ Accuracy(p) or entropy(p)\;
                \For{$i=1$ \KwTo\,$m$}{
                $\Phi_3 \leftarrow$ CC$({\bf \Phi_1, \Phi_2},\, r_1,\, r_2,\, |S_{12}|,\,|S_{13}|,\,|S_{23}|,\,\Phi_{3,1})$\;
                $B(d_2, {\bf \Phi_1, \Phi_2}, r_1,r_2)\leftarrow B(d_2, {\bf \Phi_1, \Phi_2}, r_1,r_2) \bigcup \{\Phi_3\}$}}
                \lElse{Ignore \tcp{Challenge does not exist}\DontPrintSemicolon}}
        \KwRet{$B(d_1, {\bf \Phi_1, \Phi_2}, r_1,r_2)$, $B(d_2, {\bf \Phi_1, \Phi_2}, r_1,r_2)$} 
\end{algorithm}

\begin{algorithm}[!h]
	\caption{CBDS: Create Bin first bit Different Sign: $\Phi_{1,1}\neq\Phi_{2,1}$\label{alg:2anch_CBDS}}
	\KwIn{Anchors ${\bf \Phi_1, \Phi_2}$, responses $r_1$ $r_2$, $|S_{12}|,\,|S_{13}|,\,|S_{23}|$ under $\Phi_{1,1}\neq\Phi_{2,1}$}
	\KwOut{$B(d, {\bf \Phi_1, \Phi_2}, r_1,r_2)$ of size $m$}
 	       
	$\rho_{12}=(2|S_{12}|+1)/n-1$\;
 	        \Begin{$\Phi_{3,1}=\Phi_{1,1}$\;
 	            $\rho_{13}=(2|S_{13}|)/n-1$\;
 	            $\rho_{23}=(2|S_{23}|+1)/n-1$\;
 	            $\#\text{of challenges }=  {|S_{12}|\choose K}{n-|S_{12}| \choose |S_{13}|-K-1}$\;
 	            \uIf{${\bf K}\in \mathbb{N}\quad \&\quad \#\text{of challenges }\geq 0$}{ 
                $p\leftarrow \frac{1}{2}\left[1+\frac{r_1\arcsin{\rho_{13}}+r_2\arcsin{\rho_{23}}}{\pi/2+r_1r_2\arcsin{\rho_{12}}}\right]$\;
                $d_1=$ Accuracy(p) or entropy(p)\;
                \For{$i=1$ \KwTo\,$m$}{
                $\Phi_3 \leftarrow$ CC$({\bf \Phi_1, \Phi_2},\, r_1,\, r_2,\, |S_{12}|,\,|S_{13}|,\,|S_{23}|,\,\Phi_{3,1})$\;
                $B(d_1, {\bf \Phi_1, \Phi_2}, r_1,r_2)\leftarrow B(d_1, {\bf \Phi_1, \Phi_2}, r_1,r_2) \bigcup \{\Phi_3\}$}}
                \lElse{Ignore \tcp{Challenge does not exist}\DontPrintSemicolon}}
 	        \Begin{$\Phi_{3,1}=\Phi_{2,1}$\;
 	            $\rho_{13}=(2|S_{13}|+1)/n-1$\;
 	            $\rho_{23}=(2|S_{23}|)/n-1$\;
 	            $\#\text{of challenges }= {|S_{12}| \choose K}{n-|S_{12}| \choose |S_{23}|-K-1}$\;
 	            \uIf{${\bf K}\in \mathbb{N}\quad \&\quad \#\text{of challenges }\geq 0$}{ 
                $p\leftarrow \frac{1}{2}\left[1+\frac{r_1\arcsin{\rho_{13}}+r_2\arcsin{\rho_{23}}}{\pi/2+r_1r_2\arcsin{\rho_{12}}}\right]$\;
                $d_2=$ Accuracy(p) or entropy(p)\;
                \For{$i=1$ \KwTo\,$m$}{
                $\Phi_3 \leftarrow$ CC$({\bf \Phi_1, \Phi_2},\, r_1,\, r_2,\, |S_{12}|,\,|S_{13}|,\,|S_{23}|,\,\Phi_{3,1})$\;
                $B(d_2, {\bf \Phi_1, \Phi_2}, r_1,r_2)\leftarrow B(d_2, {\bf \Phi_1, \Phi_2}, r_1,r_2) \bigcup \{\Phi_3\}$}}
                \lElse{Ignore \tcp{Challenge does not exist}\DontPrintSemicolon}}
        \KwRet{$B(d_1, {\bf \Phi_1, \Phi_2}, r_1,r_2)$, $B(d_2, {\bf \Phi_1, \Phi_2}, r_1,r_2)$} 
\end{algorithm}

 \end{appendix}

\end{document}